\documentclass[aps,prx,showpacs,amsmath,amssymb,superscriptaddress,twocolumn,longbibliography]{revtex4-2}
\usepackage[english]{babel}
\usepackage[utf8]{inputenc}
\usepackage{amsfonts}
\usepackage[T1]{fontenc}
\usepackage[pdftex]{graphicx}
\usepackage{url}
\usepackage[titletoc,title]{appendix}
\usepackage{epstopdf}
\usepackage{amsmath} 
\usepackage{amssymb}
\usepackage{dsfont}
\usepackage{braket}
\usepackage{mathtools}
\usepackage{xcolor}
\usepackage{tikz}
\usepackage[export]{adjustbox}
\usetikzlibrary{arrows.meta}
\usepackage{hyperref}
\usepackage[normalem]{ulem}
\hypersetup{
    unicode=false, 
    pdftoolbar=false, 
    pdfmenubar=true, 
    pdffitwindow=false, 
    pdfstartview={}, 
    pdftitle={...}, 
    pdfauthor={M. Ljubotina et al.}, 
    pdfsubject={}, 
    pdfcreator={}, 
    pdfproducer={}, 
    pdfkeywords={MPS, tangent spaces, time evolution}, 
    pdfnewwindow=true, 
    colorlinks=true, 
    linkcolor=black, 
    citecolor=black, 
    filecolor=black, 
    urlcolor=black 
}

\newcommand{\tr}{\mathop{\rm Tr}}
\newcommand{\id}{\text{\usefont{U}{bbold}{m}{n}1}}

\makeatletter
\newcommand{\ostar}{\mathbin{\mathpalette\make@circled\star}}
\newcommand{\make@circled}[2]{%
  \ooalign{$\m@th#1\smallbigcirc{#1}$\cr\hidewidth$\m@th#1#2$\hidewidth\cr}%
}
\newcommand{\smallbigcirc}[1]{%
  \vcenter{\hbox{\scalebox{0.77778}{$\m@th#1\bigcirc$}}}%
}
\makeatother

\newcommand{\mytitle}{Tangent space generators of matrix product states and exact Floquet quantum scars}

\begin{document}
    \title{\mytitle} 
    \author{Marko Ljubotina}
    \affiliation{IST Austria, Am Campus 1, 3400 Klosterneuburg, Austria}
    \author{Elena Petrova}
    \affiliation{IST Austria, Am Campus 1, 3400 Klosterneuburg, Austria}
    \author{Norbert Schuch}
    \affiliation{University of Vienna, Faculty of Physics, Boltzmanngasse 5, 1090 Wien, Austria}
    \affiliation{University of Vienna, Faculty of Mathematics, Oskar-Morgenstern-Platz 1, 1090 Wien, Austria}
    \author{Maksym Serbyn}
    \affiliation{IST Austria, Am Campus 1, 3400 Klosterneuburg, Austria}
    
    \begin{abstract} 
        The advancement of quantum simulators motivates the development of a theoretical framework to assist with efficient state preparation in quantum many-body systems. 
        Generally, preparing a target entangled state via unitary evolution with time-dependent couplings is a challenging task and very little is known about the existence of solutions and their properties. 
        In this work we develop a constructive approach for preparing matrix product states (MPS) via continuous unitary evolution. 
        We provide an explicit construction of the operator which exactly implements the evolution of a given MPS along a specified direction in its tangent space. 
        This operator can be written as a sum of local terms of finite range, yet it is in general non-Hermitian. 
        Relying on the explicit construction of the non-Hermitian generator of the dynamics, we demonstrate the existence of a Hermitian sequence of operators that implements the desired MPS evolution with the error which decreases exponentially with the operator range. 
        The construction is benchmarked on an explicit periodic trajectory in a translationally invariant MPS manifold. 
        We demonstrate that the Floquet unitary generating the dynamics over one period of the trajectory features an approximate MPS-like eigenstate embedded among a sea of thermalizing eigenstates. 
        These results show that our construction is useful not only for state preparation and control of many-body systems, but also provides a generic route towards Floquet scars --- periodically driven models with quasi-local generators of dynamics that have exact MPS eigenstates in their spectrum. 
    \end{abstract}
    \maketitle

    \section{Introduction}
    \label{sec:introduction}

    The ongoing progress in the development of quantum simulators~\cite{Houck:2012vz,Devoret13,DOHERTY20131,Bloch2008,lewenstein2012ultracold,Blatt2012,Labuhn2016,Bernien2017,Browaeys2020,Dolev20}, calls for the understanding of efficient control frameworks for interacting quantum systems. 
    In this context, a critical challenge is the task of state preparation which entails generating non-trivial states through unitary dynamics, employing a limited set of control parameters that are available in a specific quantum simulator. 
    In recent years, many different approaches to quantum control have been proposed, based either on analytics or numerics. 
    Analytic examples include those based on the adiabatic theorem~\cite{adiabatic}, as well as counter-diabatic methods~\cite{Demirplak:2003wf,Berry_2009,Campo12,Campo13,Opat14,Saber14,PhysRevA.96.013431,PhysRevX.4.021013,sels2017,PhysRevA.98.043436,Demirplak:2003wf,sels2017}. 
       
    Numerical approaches to quantum control often rely on brute force optimization procedures of the fidelity of the target state~\cite{CRAB,PhysRevA.95.012317,PhysRevA.84.022305,Jesper21,Jesper21-2}. 
    Calculating the fidelity resulting from a given unitary evolution is computationally costly due to the exponential growth of the Hilbert space, which scales with the number of components in interacting quantum systems. 
    In one dimension this problem can be mitigated by expressing the quantum wave function in the form of matrix product states (MPS)~\cite{Schollwock}. 
    As a consequence, MPS-based numerical algorithms~\cite{Doria11,M_ller_2022,Boscain21,PhysRevA.95.012317,PhysRevA.84.022305,Jesper21,Jesper21-2} were applied to a number of control problems in one dimension. 
    More recently, machine learning numerical approaches were also introduced~\cite{Bukov18,Zhang_2019,Niu:2019pwi,Bukov22}. 
    
    In addition to enabling efficient numerical simulations, MPS were also used in an analytic approach~\cite{Ljubotina2022} introduced by some of the authors of the present paper and their collaborators. 
    Specifically, Ref.~\cite{Ljubotina2022} provided a variational construction of the optimal controls for generating an evolution of a specific MPS along a given direction from its tangent manifold, see Fig.~\ref{fig:schematic} below. 
    The variational construction relied on the minimization of the so-called leakage, which quantifies the discrepancy between the desired unitary evolution and the one generated by the available controls from a chosen operator basis. 
    This approach may be viewed as a counterpart of the time dependent variational principle (TDVP), which constructs the optimal projection of the dynamics generated by a specific Hamiltonian onto the MPS manifold~\cite{Haegeman}.  
    While TDVP minimizes the leakage over directions in the tangent space,  Ref.~\cite{Ljubotina2022} introduced leakage minimization with respect to the generators of the unitary dynamics for a fixed tangent space direction. 
    
    The variational approach~\cite{Ljubotina2022}, although being practical, left a crucial unanswered question: what does it take to generate \emph{exact dynamics} along a given tangent space direction of the MPS manifold? 
    We address this question in the present work by providing an explicit operator construction for the generators of the MPS tangent space evolution, similar to the construction of parent Hamiltonians~\cite{Fannes1992,PerezGarcia2007}. 
    Our approach gives an exact generator of the MPS tangent space, written as a translationally invariant sum of local operators of finite range which is in general non-Hermitian. 
    While these non-Hermitian generators deserve a future study, their implementation may generally not be possible in real experiments where evolution is often unitary, which in turn requires Hermitian generators. 
    
    Hence, in the second part of our work we provide a construction for a sequence of approximate Hermitian generators of the tangent space with an increasing range of support of the local operators. 
    We demonstrate that upon increasing the support of the operators, the error quantified by leakage decreases exponentially. 
    Moreover, we show numerically that using optimization over the remaining free parameters, the sequence of approximate Hermitian operators may be made convergent in operator space, which can lead to exact Hermitian quasi-local tangent space generators. 
    While the existence of such operators is known for ground states of gapped Hamiltonians from quasi-adiabatic continuation~\cite{Hastings_2005}, our work provides a specific route to the construction of such operators that relies only on the MPS and does not make use of parent Hamiltonians.
    
    Finally, we use our approach to numerically construct the Hermitian family of generators of dynamics over an example closed-loop MPS trajectory. 
    We verify that our numerical construction becomes progressively more exact with an increasing range of support for the local terms in the generators, and consider the properties of the resulting Floquet unitary operator, which corresponds to the dynamics integrated over a period of the MPS trajectory. 
    By construction, this Floquet unitary features an MPS eigenstate that is becoming exact when increasing the support of the local operators. 
    At the same time the majority of eigenstates of the constructed Floquet unitary are chaotic. 
    Therefore, the resulting Floquet system provides an example of weak violation of the so-called Floquet eigenstate thermalization hypothesis (ETH)~\cite{D_Alessio_2013,D_Alessio_2016,Lazarides14,Kuwahara_2016,Mori_2016}. 
    Floquet ETH implies that all eigenstates of a Floquet operator correspond to infinite temperature, whereas the Floquet operator constructed from quasi-local MPS tangent space generators features an exact MPS eigenstate dubbed a Floquet quantum scar~\cite{wenwei18TDVPscar,Turner2017}.  

    The construction of the Floquet model with an ETH-violating MPS eigenstate establishes a surprising relation between the generators of tangent space dynamics and Floquet scars. 
    While scars in a Floquet setting were considered previously~\cite{Nishad21,Sugiura2019,Haldar2019,Mukherjee2020,Mizuta2020}, our results suggest that any MPS periodic trajectory has a corresponding Floquet model with quasi-local generators of dynamics which implement the dynamics along the MPS trajectory exactly. 
    Thus, our work suggests that scars in a Floquet setting may be quite abundant, and calls for a systematic exploration of Floquet scars and their use as a means of quantum control in many-body systems.

    The remainder of the paper is structured as follows. 
    In Sec.~\ref{sec:control_mps} we define the quantum control problem over an MPS manifold and introduce the necessary notations. 
    Then, in Sec.~\ref{sec:nonhermitian} we present the construction of the exact non-Hermitian driving operators. 
    This is followed by Sec.~\ref{sec:hermitian_operator}, where we construct the sequence of approximate Hermitian driving operators, and bound the leakage stemming from this approximation.
    Having discussed the theoretical aspect, we proceed, in Sec.~\ref{sec:example}, by demonstrating our construction and its performance for a specific MPS trajectory. 
    Furthermore, we show that the operator series can be made convergent to an exact seemingly quasi-local operator through the use of the free parameters.
    Finally, we discuss some of the remaining open questions and potential future directions in Sec.~\ref{sec:discussion}. 
    The paper is concluded by several Appendices that discuss the details of the free parameters present in the construction of the generators, the derivation of the leakage bound, discuss the MPS trajectory studied in this work, and show further numerical details concerning the Floquet propagator we obtain with our approach. 

    \section{MPS control problem and notations}
    \label{sec:control_mps}

    In this section we first provide a general formulation of the control problem of MPS as considered in this work.
    We then introduce the relevant properties of MPS that will be used throughout the work. 

   \begin{figure}[t]
        \centering
        \includegraphics[width=.99\linewidth]{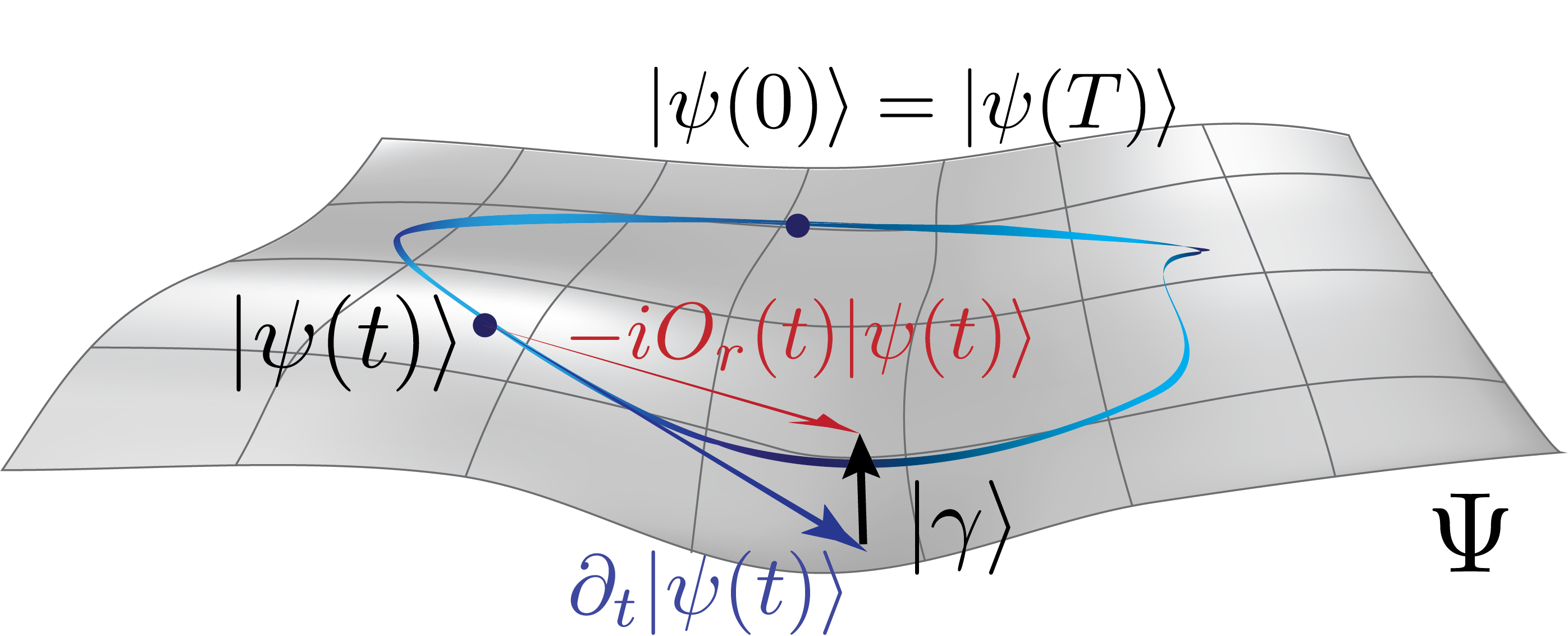}
        \caption{
            Schematic representation of the desired trajectory $|\psi(t)\rangle$, here chosen to be closed with $|\psi(0)\rangle=|\psi(\tau)\rangle$, within the MPS manifold $\Psi$. 
            We seek to identify an operator $O_r$ generating the desired dynamics at each point along the trajectory, such that the discrepancy between the desired direction in the tangent space and the unitary evolution is vanishing, $|\gamma\rangle=0$.
        }
        \label{fig:schematic}
    \end{figure}

    \subsection{Finding generators of a MPS tangent space}
    \label{sec:control_mps_sub_control}

    We start by defining the problem of finding generators of dynamics along a certain direction within an MPS manifold. 
    Let us consider a one-dimensional chain of $N$ spins $s$, we can then define a Hilbert space of all states $\mathcal{H}=\mathbb{C}^{\otimes d^N}$, where $d=2s+1$ is the local Hilbert space dimension associated with a single site. 
    We can now define a trajectory of quantum states $|\psi(t)\rangle$, with $t\in[0,\tau]$, within an MPS manifold $\Psi\subseteq\mathcal{H}$, as sketched in Fig.~\ref{fig:schematic}. 
    While the details of MPS will be defined below, we are ready to formulate the control problem. 
    Specifically, given a state $|\psi(t)\rangle$ and a desired evolution direction $\partial_t|\psi(t)\rangle$, we aim to find an operator $O(t)$ which generates the evolution
    \begin{equation}
        \partial_t|\psi(t)\rangle=-iO(t)|\psi(t)\rangle.
        \label{eq:schrodinger}
    \end{equation}
    We note that this operator is in general dependent on the point along the trajectory. 
    Figure~\ref{fig:schematic} illustrates a schematic of this setup, introducing the vector $|\gamma\rangle$, given by the difference of the right and left hand sides of Eq.~\eqref{eq:schrodinger}. 
    This vector vanishes if an exact solution is found and its norm, known as leakage, will be used to quantify the quality of the solution. 
    
    We will be interested in the solution of the above problem when the operator $O$ (for brevity we omit the time dependence in $O$ and $|\psi\rangle$ from this point on) is written as a sum of $r$-local terms, denoted as $O_r$,
    \begin{equation}
        O_r=\sum_i o_{i,i+r-1}.
        \label{eq:locality}
    \end{equation}
    Here $o_{i,i+r-1}$ is a combination of local operators acting non-trivially on sites $[i,i+r-1]$ and the sum is taken over the entire chain.
    While numerically we use periodic boundary conditions, our analytic construction will consider an infinite system, where boundary conditions are not relevant. 
    Before we can present the explicit construction of $O_r$ we need to introduce some notations of MPS and discuss the properties of MPS used in our construction. 
    
    \subsection{Properties of MPS and their tangent spaces}
    \label{sec:control_mps_sub_mps}

    First, we define an MPS with bond dimension $\chi$ that is used to parametrize quantum states throughout this work as  
    \begin{equation}
        |\psi\rangle=\sum_{\underline{s}}
        \tr\left[{\prod_{i=1}^NA_i^{s_i}}\right]|\underline{s}\rangle,
        \label{eq:mps_def}
    \end{equation}
    where $A_i^{s_i}$ are $\chi\times\chi$ complex matrices. 
    In this work we will consider the MPS to be translationally invariant, with all matrices being identical, $A_i^s=A^s$. 
    However, we note that our construction can also be extended to non-translationally-invariant MPS, however, all derivations here will be performed for the translationally invariant case. 
  
    We can now define the one-site transfer matrix
    \begin{equation}
        T=\sum_{s}A^s\otimes \overline{A}^s
        \label{eq:transfer_matrix}
    \end{equation}
    where $\overline{A}^s$ is the complex conjugate of $A^s$. 
    Considering the eigenvalues and eigenvectors of the transfer matrix, we define the left and right eigenvectors as $T|R_i)=\lambda_i|R_i)$ and $(L_i|T=\lambda_i(L_i|$, which satisfy the condition $(L_i|R_j)=\delta_{i,j}$ for all $i,j$. 
    We assume both that the state is normalized and that the transfer matrix has a unique dominant eigenvalue, with $\lambda_1=1$, which always holds for injective MPS, which we will discuss below. 
    We then introduce a simplified notation for the corresponding dominant eigenvectors $|R_1)=|R)$ and $(L_1|=(L|$. 
    These become particularly relevant when one considers the $r$-site transfer matrix, $T^r$. 
    When the number of sites $r$ is large, $r\to\infty$, the transfer matrix converges to $T^\infty\equiv\lim_{r\to\infty}T^r=|R)(L|$. 
    
    Furthermore, in order to describe the time evolution we will need to consider the tangent space $\mathcal{T}_{|\psi\rangle}\Psi$ of the MPS manifold $\Psi$ at an arbitrary point $|\psi\rangle$. 
    We introduce the notation $\partial A\equiv\partial_t A(t)$, where $\partial_t A(t)$ is defined with the specific choice of the MPS trajectory.
    Using the fact that $\partial_t|\psi\rangle$ is an element of the tangent space $\mathcal{T}_{|\psi\rangle}\Psi$, we note that $\partial_t|\psi\rangle$ must be orthogonal to the original MPS $\psi$ in the thermodynamic limit,
    \begin{equation}
        \langle\psi|\partial_t|\psi\rangle=N(L|\sum^d_{\sigma=1}(\partial_tA^\sigma(t))\otimes\overline{A^\sigma(t)}|R)=0.
        \label{eq:orthogonality}
    \end{equation}
    Note that in general this overlap can take any purely imaginary value, however, we can always set the value to zero without loss of generality as this amounts to a simple modification of the global phase of our state. 

\begin{figure}[t]
        \centering
        \includegraphics[width=.7\linewidth]{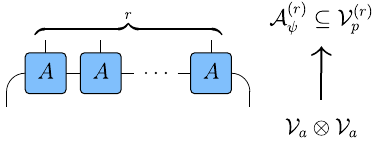}
        \caption{
            We can view a section of the MPS containing $r$ sites as a map from the doubled auxiliary space $\mathcal{V}_a\otimes\mathcal{V}_a$ to a subspace $\mathcal{A}_\psi^{(r)}$ of the Hilbert space of $r$ sites ($\mathcal{V}_p^{(r)}=\mathbb{C}^{d^r}$). 
            In this picture the injectivity of the MPS implies $\dim\mathcal{A}_\psi^{(r)}=\chi^2$ for sufficiently large $r$. 
        }
        \label{fig:mps_spaces}
    \end{figure}

    An important restriction for our construction of the tangent space generators is that the MPS is injective~\cite{Fannes1992,PerezGarcia2007}. 
    This can be understood by considering the map between the doubled space of virtual (bond) indices $\mathcal{V}_a\otimes\mathcal{V}_a=\mathbb{C}^{\chi^2}$ of dimension $\chi^2$ and the physical states in the Hilbert space $\mathcal{A}_\psi^{(r)}\subseteq\mathcal{V}_p^{(r)}=\mathbb{C}^{d^r}$ of $r$ sites of dimension $d^r$, where $d$ is the local Hilbert space dimension. 
    Such a map is illustrated in Fig.~\ref{fig:mps_spaces}.
    Provided that $r$ is sufficiently large (in generic situations $d^r\ge\chi^2$, though this will in general only be a lower bound on the minimal $r$), injectivity requires that the rank of the map must be equal to $\chi^2$, or equivalently $\dim\mathcal{A}_\psi^{(r)}=\chi^2$. 
    Here $\mathcal{A}_\psi^{(r)}$ can be viewed as the subspace of the total Hilbert space of $r$ sites, representable by the MPS on $r$ sites. 
    Finally, we will denote the orthogonal complement of $\mathcal{A}_\psi^{(r)}$ in  $\mathcal{V}_p^{(r)}$ by ${\mathcal{B}_\psi^{(r)}}$, i.e., 
    $\mathcal{V}_p^{(r)}=\mathcal{A}_\psi^{(r)}\oplus {\mathcal{B}_\psi^{(r)}}$; it contains the $r$-site states which are orthogonal to the ones representable by the MPS.
    Similarly, we can define $\mathcal{A}_{\partial_j\psi}^{(r)}\subseteq\mathcal{V}_p^{(r)}$ as the subspace representable by the derivative of the MPS, with the derivative taken on the $j$-th site (the corresponding map can be obtained by replacing the $j$-th tensor $A$ by $\partial A$ in Fig.~\ref{fig:mps_spaces}). 
    These spaces will play an important role later in the definitions of the free parameters of our generators. 

    Injectivity has important implications for the MPS. 
    First, injectivity implies the uniqueness of the largest dominant eigenvalue $\lambda_1=1$ (set to one by normalization) of the transfer matrix defined above~\cite{PerezGarcia2007,Zauner_2015,Haegeman_2017}.
    Second, injectivity also allows us to define a left-inverse $I_r$ of the map illustrated in Fig.~\ref{fig:mps_spaces} according to the condition
    \begin{equation}
        \includegraphics[width=.65\linewidth]{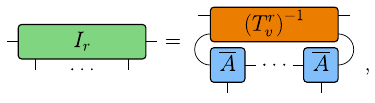}
        \label{eq:inverseDef}
    \end{equation}
    where $T_v^r$ is transfer matrix of $r$-sites, $T^r$, whose legs are then reordered to form a matrix in the vertical direction by merging the two bottom and two top legs as shown in the equation.
    Such a left-inverse is unique and annihilates all states in $\mathcal{B}_\psi^{(r)}$ while acting as a regular inverse on the remaining subspace $\mathcal{A}_\psi^{(r)}$. 
    The action of such a left-inverse operator when applied to an MPS state can be represented as
    \begin{equation}
        \includegraphics[width=.79\linewidth]{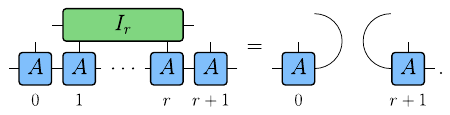}
        \label{eq:inverse}
    \end{equation}
    This operator will play a key role in the construction of the tangent space generators below. 

    \section{Construction of tangent space generators}
    \label{sec:generators}
    
    This section presents the main details of our construction of the tangent space generators. 
    In Section~\ref{sec:nonhermitian} we present the explicit form of a finite range non-Hermitian tangent space generator and discuss the available free parameters that allow us to modify the operator. 
    However, since generators of unitary dynamics must be Hermitian, we also demonstrate a construction of a sequence of approximate Hermitian generators of the tangent space dynamics in Sec.~\ref{sec:hermitian_operator}. 
    Moreover, we show that the leakage that controls the quality of the approximation decreases at least exponentially with support $r$ for this sequence of Hermitian operators. 
    
    \subsection{Non-Hermitian generators}
    \label{sec:nonhermitian}   

\begin{figure}[t]
        \centering
        \includegraphics[width=.99\linewidth]{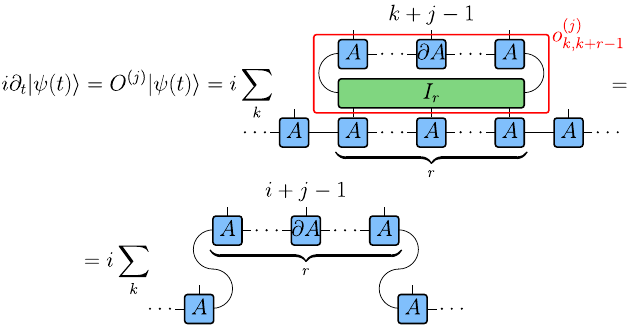}
        \caption{Schematic construction of the tangent space generator. Note that the position of the derivative term $j$ is arbitrary, in fact one can even take a sum of contributions with different positions (see text for details).\label{fig:mps_operator_construction}}
    \end{figure}

    The basic idea behind the construction of the driving operator $O_r$ is presented in Fig.~\ref{fig:mps_operator_construction}. 
    By applying the left-inverse $I_r$ to $r$ contiguous tensors in the MPS and using Eq.~\eqref{eq:inverse} we can effectively replace the tensors in this region. 
    Thus, the solution to the tangent space generators is given by the sum of local range-$r$ operators  
    \begin{equation}
        \label{eq:o-answer}
        \includegraphics[width=0.65\linewidth]{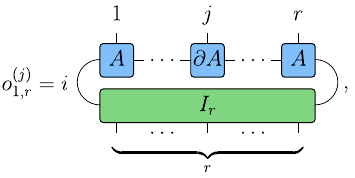}
    \end{equation}
    such that $O^{(j)}_r=\sum_k o_{k,k+r-1}^{(j)}$. 
    Here $j$, $1\leq j \leq r$, represents the site at which the derivative term is applied. 
    This is an explicit solution that provides the generators of the MPS tangent space, and it is guaranteed to exist as soon as the range of the operators $r$ is sufficiently large, such that the left-inverse $I_r$ can be found. 
    However, this solution is generally non-Hermitian -- an issue that we will address in detail in the next section. 
    Furthermore, this solution is by no means unique, and below we discuss two qualitatively different sets of free parameters, that allow to change the form of operator $O_r$, while maintaining the property that it solves Eq.~\eqref{eq:schrodinger} exactly~\footnote{See Appendix \ref{app:free_params} for a detailed discussion of the free parameters.}. 

    The first set of free parameters originates from the ``redistribution'' of the derivative term among $r$ sites within the range of operator $I_r$. 
    Note that in Eq.~\eqref{eq:o-answer} we position the derivative term $\partial A_j$ at some position $j$. 
    However, this position is arbitrary, and in fact we are free to choose any linear combination of derivatives 
    at sites $j=1,\dots,r$ with (not necessarily positive) weights $\alpha_j$ such that $ \sum_{j=1}^{r}\alpha_j=1$. 
    Such spreading of the derivative leads to a modified local operator that is expressed as a linear sum
    \begin{equation}
        o_{i,i+r-1}=\sum_{j=1}^{r}\alpha_jo_{i,i+r-1}^{(j)}, 
        \label{eq:alpha}
    \end{equation}
    with operators $o_{i,i+r-1}^{(j)}$ defined in Eq.~\eqref{eq:o-answer}.
    A similar transformation then also follows for the translationally invariant sum $O_r=\sum_j\alpha_jO_r^{(j)}$. 
    As we will see later, the redistribution of the derivative among $r$ sites can have a non-trivial effect on the resulting tangent space generator. 

    The second set of free parameters emerges from the fact that the operator $o_{1,r}$ is defined through the left-inverse in Eq.~\eqref{eq:o-answer}. 
    From the properties of the left-inverse discussed in Sec.~\ref{sec:control_mps_sub_mps} we observe that $o_{1,r}$ is a map between the subspaces $\mathcal{A}_\psi^{(r)}\to\mathcal{A}_{\partial_j\psi}^{(r)}$, and annihilates states in the complement space, $\mathcal{B}_\psi^{(r)}$. 
    Since the operator $o_{1,r}$ acts non-trivially only on a subspace of the full physical Hilbert space $\mathcal{V}_p^{(r)}$ of $r$ sites, we can modify it as 
    \begin{equation}
        o'_{1,r}=o_{1,r}+\sum_{x,y}c_{x,y}|x\rangle\langle y|,
        \label{eq:free_params}
    \end{equation}
    where $x\in\{\mathcal{V}_p^{(r)}\}$ and $y\in\{{\mathcal{B}_\psi^{(r)}}\}$ run over the basis of their respective spaces. 
    Physically this modification corresponds to adding a (non-Hermitian) local term to the operator $o_{1,r}$ which annihilates $|\psi\rangle$, similar to the construction of the parent Hamiltonian~\cite{Giudici_2022}. 
    As such a local term acts non-trivially only on the complement space $\mathcal{B}_\psi^{(r)}$, it annihilates the MPS state and thus the modified operator $o'_{1,r}$ still solves $-i\sum_io_{i,i+r-1}'|\psi\rangle=\partial_t|\psi\rangle$.
    We note that in what follows we restrict the above freedom to $x\in\{{\mathcal{B}_\psi^{(r)}}\}$ rather than the complete Hilbert space of $r$ sites. 
    This restriction is important since the construction of the Hermitian generators will require the minimization of the action of $o'^\dagger_{1,r}$ on the MPS state $|\psi\rangle$. 
    The redefinition of $o_{1,r}$ according to Eq.~\eqref{eq:free_params}, provided $x\in\{{\mathcal{B}_\psi^{(r)}}\}$, will thus ensure that the free parameters do not affect the action of ${o'}^\dagger_{1,r}$ on the MPS state $|\psi\rangle$.
    As a result we will be left with $(d^r-\chi^2)^2$ free real parameters $c_{x,y}$. 

    \subsection{Approximate Hermitian driving operators and the leakage bound}
    \label{sec:hermitian_operator}
    
    As we discussed in the previous section, the solution to the generators of the MPS tangent space is not unique. 
    In particular, any operator $O_r$ which solves Eq.~\eqref{eq:schrodinger} can be modified as $O_r+X$, and will remain a good solution as long as the operator $X$, which is not necessary Hermitian, annihilates the MPS state $X|\psi\rangle=0$. 
    This freedom is at the core of our construction of the Hermitian driving operator below. 
    In particular, we notice that the operator $O_r^\dagger$, while not satisfying the property $O_r^\dagger|\psi\rangle=0$ exactly, satisfies it progressively better with increasing $r$, as we show below. 
    This allows us to take the following construction of the Hermitian generators of the tangent space, 
    \begin{equation}
        H_r=O_r+O_r^\dagger,
        \label{eq:hermitian}
    \end{equation}
    which leads to an exponentially decreasing leakage with increasing operator support $r$.
    
    Specifically, in Appendix~\ref{app:bound} we demonstrate that the following bound on the norm of the state $O_r^\dagger|\psi\rangle$ holds for sufficiently large ranges $r$:
    \begin{equation}
        \lVert O^\dagger_r|\psi\rangle\rVert_2\le Nc|\lambda_2|^{r}\, \left(\frac{r-1}{2}\right)^{\chi^2-1}.
        \label{eq:bound}
    \end{equation}
    Here $c\in\mathbb{R}$ is a scalar constant dependent on various parameters of the MPS and the rate of its change but independent of $r$~\footnote{See Appendix \ref{app:bound} for details on the constant and the derivation of the bound.}, $\lambda_2$ is the subleading eigenvalue (i.e. second largest in absolute value) of the transfer matrix $T$ in the usual horizontal direction and $\lVert \cdot\rVert_2$ is the vector $2$-norm. 
    Note that injectivity of the MPS guarantees that $|\lambda_2|<1$, thus implying an exponential decay of the norm of $\lVert O_r^\dagger|\psi\rangle\rVert_2$ as $\mu^r$ for any positive $\mu$ satisfying $|\lambda_2|<\mu<1$.

    This bound can then be used to bound the leakage from the evolution generated by these operators. 
    As shown in Fig.~\ref{fig:schematic}, we define the leakage vector as the difference between the desired and actual change of state
    \begin{equation}
        |\gamma\rangle=-iH_r|\psi\rangle-\partial_t|\psi\rangle.
        \label{eq:leakage}
    \end{equation}
    Since the non-Hermitian operator $O_r$ generates the exact dynamics $-iO_r|\psi\rangle=\partial_t|\psi\rangle$ we can quickly identify $|\gamma\rangle=-iO_r^\dagger|\psi\rangle$.
    From this it follows that the norm of the leakage is directly bounded by the bound from Eq.~\eqref{eq:bound}. 

    We have demonstrated that the sequence of Hermitian operators $H_r$ satisfies Eq.~\eqref{eq:schrodinger} with exponentially increasing precision as we increase the support~$r$. 
    At the same time, operators with different supports, for instance $H_r$ and $H_{r+1}$, are not guaranteed to be close to each other in terms of operator norms. 
    Such convergence in operator space with increasing $r$ is a desirable property, since it would allow one to define a \emph{quasi-local} Hermitian generator of the tangent space. 
    While demonstrating such a property analytically remains an interesting direction for future work, we will demonstrate in the following section that it is possible to achieve quasi-locality of $H_r$ using numerical optimization over the free parameters in the redefinition of the operator $o'_{1,r}$ in Eq.~\eqref{eq:free_params}.

    \begin{figure*}[t!]
        \centering
        \includegraphics[width=0.95\textwidth]{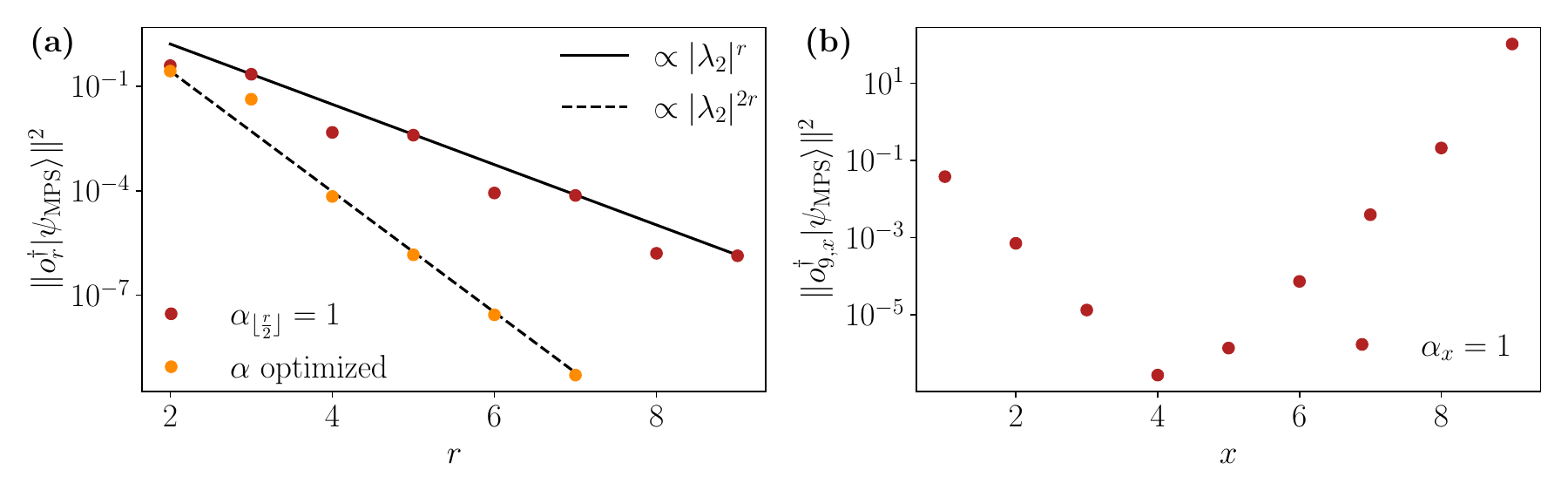}
        \caption{
            (a) The norm of $o^\dagger|\psi\rangle$ shows a clear exponential decay in agreement with the predictions from the upper bound. 
            The dark red points correspond to placing the derivative in the middle of the $r$ sites while the orange points are obtained by minimizing the norm over $\alpha_j$. 
            Optimizing over the derivative distribution $\alpha_j$ will typically lead to faster convergence. 
            The lines show exponential fits to the data, note that the exponent of the solid line is precisely the one predicted by the bound in Eq.~\eqref{eq:bound}, whereas after optimization over $\alpha_i$ the decay is faster. 
            (b) The choice of $\alpha_j$ can have a strong effect on the norm which is exponentially sensitive to the location of the term where we insert the derivative.
            Here we only show the simple cases where the derivative is fully localized on a given site $x$ and all other $\alpha_y=0;~\forall y\neq x$. 
        }
        \label{fig:exp_convergence_alpha}
    \end{figure*}

    \section{From local generators of the MPS tangent space to a Floquet model}
    \label{sec:example}
    
    In this section we implement the construction of Hermitian tangent space generators introduced in the previous section. 
    We start with a definition of an example MPS trajectory in Sec.~\ref{sec:mps-def}. 
    Next, in Sec.~\ref{sec:example_convergence} we numerically demonstrate the bound from Sec.~\ref{sec:hermitian_operator} which implies an exponentially decreasing leakage with $r$ and show that the sequence of operators $H_r$ can be made quasi-local in the limit $r\to\infty$. 
    Finally, in Sec.~\ref{sec:num_floquet} we turn to study the unitary dynamics over the entire period of the chosen closed MPS trajectory. 
    We show that the resulting Floquet unitary has an MPS-like eigenstate which we dub a Floquet quantum scar.
    
    \subsection{MPS manifold and trajectory}
    \label{sec:mps-def}
    
    We will use the MPS trajectory found in Ref.~\cite{Michailidis2020} from the TDVP projection of unitary Hamiltonian dynamics onto a low bond dimension MPS manifold. 
    This trajectory was also used in Ref.~\cite{Ljubotina2022} to construct the generators of MPS dynamics with a variational approach. 
    The trajectory in the MPS manifold is specified by the explicit form of the $A^s$ matrices for a spin-$1/2$ system,
    \begin{equation}
        \label{eq:isingmps}
        A^s=\begin{pmatrix}
            \cos\mathfrak{d}\cos\mathfrak{b} e^{i\mathfrak{a}/2} \delta_{s,\uparrow} & \cos\mathfrak{d}\sin\mathfrak{b} e^{-i\mathfrak{a}/2} \delta_{s,\uparrow} \\
            \sin\mathfrak{d}\sin\mathfrak{b} e^{i(\mathfrak{c}-\mathfrak{a}/2)}\delta_{s,\downarrow} & \sin\mathfrak{d}\cos\mathfrak{b} e^{i(\mathfrak{c}+\mathfrak{a}/2)} \delta_{s,\downarrow}
        \end{pmatrix},
    \end{equation}
    with real MPS parameters $\mathfrak{a},\mathfrak{b},\mathfrak{c},\mathfrak{d}\in[-\pi,\pi]$. 
    The trajectory is now described by the 4 real valued periodic functions $\mathfrak{a}(t),\mathfrak{b}(t),\mathfrak{c}(t),\mathfrak{d}(t)$ with period $T\approx2.098$ shown in Fig.~\ref{fig:mps_params} in the Appendix~\footnote{The periodic functions describing the trajectory can be found in the Supplemental Material of Ref.~\cite{Michailidis2020}}.
    We note that the MPS above is not in canonical form, however, it is injective almost everywhere on the MPS manifold, and it is injective everywhere on the chosen MPS trajectory. 

    \subsection{Exponential decay of infidelity and quasi-locality}
    \label{sec:example_convergence}

    We will now study the properties of the approximate Hermitian operators in this example.
    Specifically, we will numerically evaluate the leakage as a function of the operator support $r$ and show that it decreases exponentially, as predicted by the bound in Eq.~\eqref{eq:bound}. 
    Furthermore, we will show how one can use the free parameters $c_{x,y}$, defined in Eq.~\eqref{eq:free_params}, to numerically obtain a series of operators converging to an increasingly exact quasi-local Hermitian driving operator. 

    We first look at the operators at a random point and direction within the MPS manifold, specifically we take $(\mathfrak{a},\mathfrak{b},\mathfrak{c},\mathfrak{d})=(1.05, -0.48, 0.39, 1.2)$ and $\partial_t(\mathfrak{a},\mathfrak{b},\mathfrak{c},\mathfrak{d})=(-3.81, 1.29, 2.1, -0.49)$. 
    In Fig.~\ref{fig:exp_convergence_alpha}(a) we show the exponential decay of $\lVert o_r^\dagger|\psi\rangle\rVert^2_2$ with $r$, which indicates an exponential decrease in leakage for the Hermitian operator~\eqref{eq:hermitian}, as predicted by the bound. 
    Note that the two sets of points shown in Fig.~\ref{fig:exp_convergence_alpha}(a) differ only in the choice of the distribution of the derivative term over the $r$ sites ($\alpha_j$), with the faster converging set being optimized over $\alpha_j$ while in the other case we simply put the derivative term on the middle site. 
    We note that this simple choice is in general a safe choice when restricting the derivative term to a single site, as shown in Fig.~\ref{fig:exp_convergence_alpha}(b) for a random state in our MPS manifold, though naturally optimization can provide better results.
    Importantly, this optimization is a simple positive semi-definite quadratic problem with $r-1$ variables and thus not computationally intensive. 
    
    After numerically demonstrating the exponential decrease of the leakage with operator range, which is also predicted by the bound~\eqref{eq:bound} proven in Appendix~\ref{app:bound}, we consider the operator space convergence of the driving operators. 
    As noted at the end of Sec.~\ref{sec:nonhermitian} the free parameters $c_{x,y}$ (defined in Eq.~\eqref{eq:free_params}) are now restricted further as we do not wish the free parameters to lead to any additional action in either $O^\dagger_r|\psi\rangle$ or $O_r|\psi\rangle$. 
    If we can achieve exponential convergence in operator space, this would imply that exact driving can be achieved with a quasi-local Hermitian operator. 
    Specifically, we shall consider the Frobenius norm $\lVert H_r\rVert_F$, due to its invariance under unitary operations. 
    This allows us to expand $H_r=\frac{1}{2^r}\sum_{\underline{s}}c_{\underline s}\otimes_{i=1}^{r}\sigma^{s_i}$ over a Pauli basis, where $\underline{s}$ is a string of length $r$ with entries $s_i\in\{x,y,z,0\}$ representing the three Pauli matrices and the identity matrix respectively. 
    We can then express the norm as $\lVert H_r\rVert_F=\sum_{\underline s}c_{\underline s}^2$. 

    This approach allows us to elegantly handle translational invariance. 
    Recall that our construction gives us access to $H_r$ as a sum of local densities $h_{i,i+r-1}=o_{i,i+r-1}+o_{i,i+r-1}^\dagger$. 
    Clearly when expanding this over a Pauli basis there is some freedom as to which $h_r$ the elements with supports smaller than $r$ are allocated to. 
    In the Pauli basis we can resolve this ambiguity by considering only strings $\underline s$ where the first element is not an identity matrix. 
    Any elements where this is not the case are shifted to the left and added to the corresponding element.
    Thus entries do not repeat for different values of $i$ and translational invariance can be taken into account by computing an upper bound $\lVert H_r\rVert_F\le N\lVert h_r\rVert_F$. 
    To attempt to achieve convergence in the operators as $r$ increases, we will determine the free parameters $c_{x,y}$ such that the norm $\lVert h_{r-1}-h_r\rVert_F$ is minimized. 

\begin{figure}[t]
        \centering
        \includegraphics[width=1\linewidth]{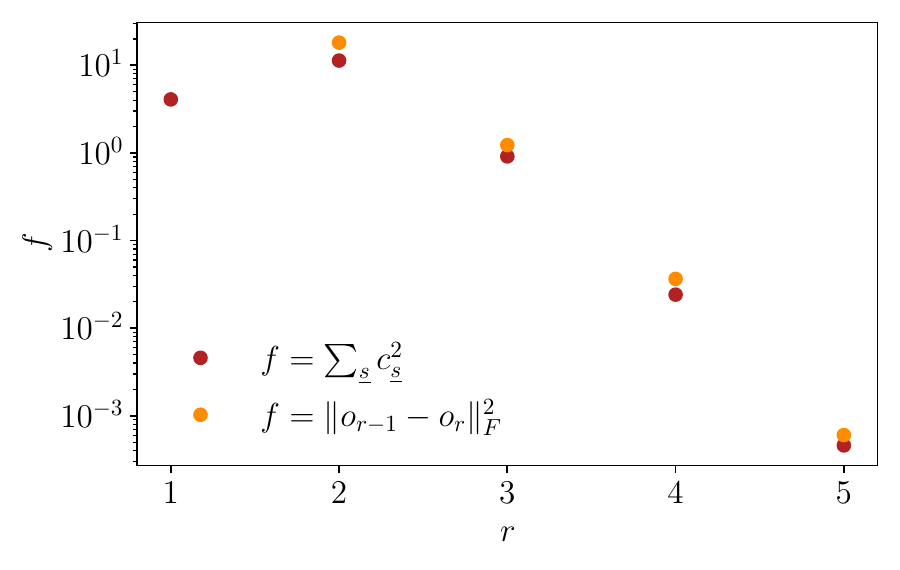}
        \caption{
            Operator space convergence properties after optimization over the free parameters aimed at obtaining a convergent operator series with respect to the Frobenius operator norm. 
            The dark red points show the weight of the operator corresponding to Pauli strings with support $r$ for $o_5$.
            The orange points show the norm of the difference between solutions with consecutive supports. 
        }
        \label{fig:operator_space_converngence}
    \end{figure}

    In Fig.~\ref{fig:operator_space_converngence} we show the convergent series and the structure of the largest operator after optimization over the free parameters $c_{x,y}$, which shows exponentially decaying weights for longer range terms.
    Our numerical exploration suggests that this convergence can be achieved for any combination of MPS parameters, although the convergence rate varies depending on the choice of MPS parameters and their derivatives. 

    \subsection{Approximate Floquet scars}
    \label{sec:num_floquet}

    Finally, we apply our approach to drive the state along a trajectory within the MPS manifold specified in Sec.~\ref{sec:mps-def}. 
    We test our construction using exact diagonalization (ED) to simulate the full quantum dynamics of the model. 
    Note that for simplicity we omit the optimization over the free parameters $c_{x,y}$. 
    We also restrict our numerical calculations to periodic boundary conditions (PBC) as that is more in line with the expected behavior in the thermodynamic limit, which we considered analytically.
    Furthermore, this allows us to resolve translational invariance which will simplify our calculations and allow us to reach larger system sizes with reasonable accuracy. 
    
    We first consider how well the constructed tangent space generators implement the dynamics along the MPS trajectory. 
    To this end, we construct the propagator which encodes the dynamics from time $0$ to $t$, 
    \begin{equation}
        U_r(t)={\mathcal T}e^{-i\int_{0}^{t}H_r(s)\text{d}s},
        \label{eq:floquet}
    \end{equation}
    where $H_r(s)$ is the approximate Hermitian tangent space generator of dynamics along the MPS trajectory constructed from the sum of range-$r$ local operators discussed in Sec.~\ref{sec:generators}. 
    Setting the time equal to the period of the trajectory, $t=\tau$, we would get the Floquet unitary, $U^{F}_r=U_r(\tau)$~\footnote{See Appendix~\ref{app:numerics} for details on the numerics. }.
  
\begin{figure}[tb]
        \centering
        \includegraphics[width=1\linewidth]{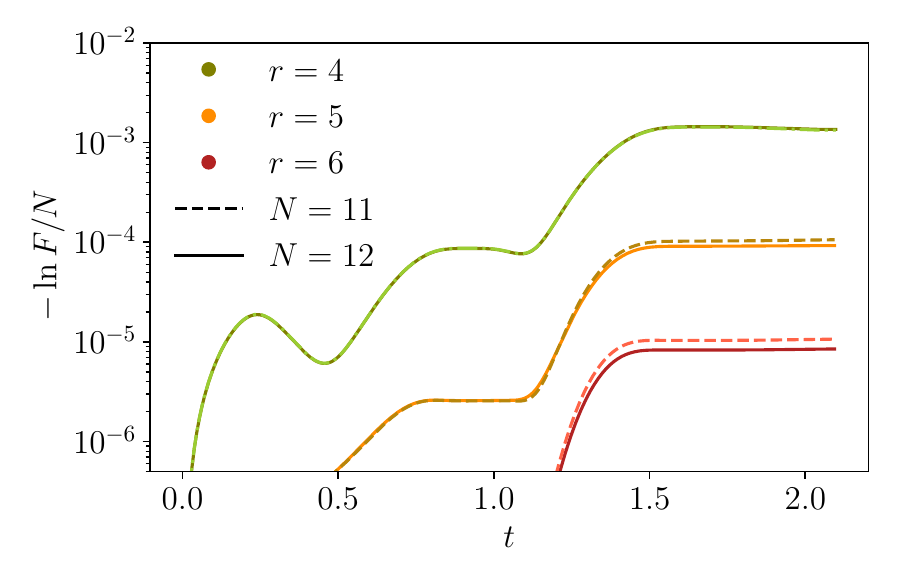}
        \caption{
            We show the log-fidelity per site defined as $-\ln F/N$ where $F=|\braket{\psi(t)|U_r(t)|\psi(0)}|^2$ is the fidelity. 
            Increasing the support $r$ shows the predicted exponential decrease in infidelity $1-F$. 
            We show data for several system sizes indicating reasonably good convergence, note that one expects the results to converge to a fixed curve for $N\to\infty$. }
        \label{fig:numerics_fidelity}
    \end{figure}

    To quantify the accuracy of our generated dynamics Fig.~\ref{fig:numerics_fidelity} shows the log-fidelity per site $f=-\ln F/N$, with $F=|\braket{\psi(t)|U_r(t)|\psi(0)}|^2$.
    We observe two distinct trends. 
    Firstly, $f$ is approximately converged with respect to the system size, suggesting that the log-fidelity per spin is a well-defined quantity. 
    Secondly, we see that increasing the range $r$ in $U_r(t)$ leads to roughly exponential improvements, as expected from the bound obtained in Sec.~\ref{sec:generators}. 

    \begin{figure}[t!]
        \centering
        \includegraphics[width=1\linewidth]{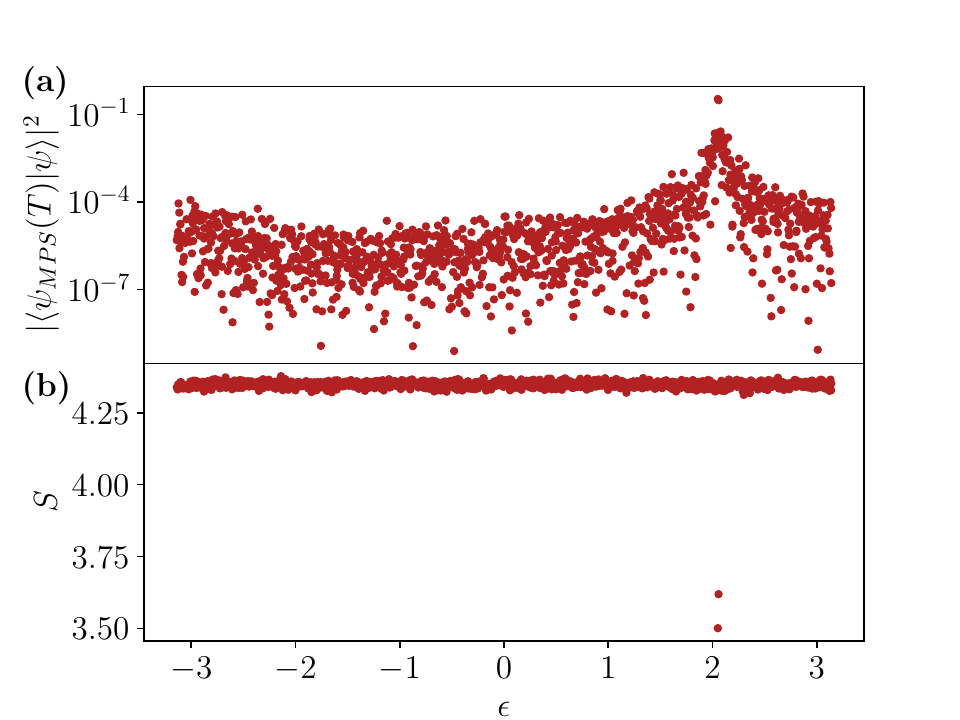}
        \caption{
            The overlap of the eigenstates of the Floquet propagator with the MPS initial state (a) as well as the entanglement of these eigenstates (b) both reveal that while nearly all eigenstates are typical, there is a single exception in the eigenstate with the largest overlap with the MPS state.
            Thus, while the propagator is otherwise chaotic, there exists a single weakly-entangled eigenstate, which can be viewed as a Floquet scar in the model. 
            Data obtained from calculations with $r=4$ and $N=14$ in the zero momentum sector with periodic boundary conditions.
        }
        \label{fig:floquet}
    \end{figure}

    Next, we study the properties of the spectrum of the Floquet unitary, $U_F$, which encodes the dynamics over one period of the MPS trajectory. 
    We observe that the density of states is smooth, with only small variations, suggesting that the system does not have any quasi-conserved quantities that would manifest as mini-bands. 
    We also find that the statistics of eigenvalues of $U_F$ show level repulsion~\cite{Note2}. 
    This suggests that the constructed Floquet unitary is far from any prethermal regimes~\cite{Ho_2023} and is instead fully chaotic. 
    At the same time, a more detailed look at the spectrum in Fig.~\ref{fig:floquet} reveals the existence of anomalous eigenstates in the spectrum.  
    
    In particular, while the majority of eigenstates of the $U_F$ seem to satisfy the Floquet eigenstate thermalization hypothesis~\cite{Lazarides14,DAlessio2014,Kim_ETH,Mori2018} in the sense of having large entanglement entropy and small overlap with weakly entangled MPS states, Fig.~\ref{fig:floquet} clearly reveals the existence of a special MPS-like eigenstate in the spectrum of $U_F$. 
    We note that in the limit $\lim_{r\to\infty}\lim_{N\to\infty}$ our construction of Hermitian tangent space generators becomes exact, and thus is expected to give at least one
    non-ergodic eigenstate of $U_F$, corresponding to the initial MPS state. 
    Numerical data supports the conclusion that the remainder of the spectrum is fully ergodic. 
    At the same time, at finite $r$, the MPS is not an exact eigenstate of the $U_F$ and its hybridization with nearby eigenstates leads to several eigenstates with anomalous entanglement and enhanced overlap with the MPS state. 
    Further results supporting these conclusions are presented in Appendix~\ref{app:numerics}. 

    Finally, our construction can, as demonstrated by the results above, also be seen as a framework that can produce perfect Floquet scars from arbitrary MPS trajectories when considering quasi-local operators. 
    For local operators with fixed supports our construction still leads to Floquet operators with one or several non-ergodic states in an otherwise fully chaotic spectrum. 
    
    \section{Discussion}
    \label{sec:discussion}

    In this work we presented a constructive approach for finding generators of dynamics in the tangent space of an MPS manifold. 
    We defined the exact tangent space generators which are generally non-Hermitian and thus do not correspond to unitary dynamics. 
    In addition, we demonstrated that tangent space generators can be made Hermitian, at the expense of introducing a small error which decreases exponentially with the range of the local terms within the operators. 
    Furthermore, we conjectured and demonstrated numerically that such a sequence of Hermitian operators can be made convergent, leading to \emph{exact quasi-local} Hermitian generators of the MPS tangent space. 

    Considering the dynamics induced by the generators of the MPS tangent space over the periodic trajectory within the MPS manifold, we defined a periodically driven Floquet model with a local Hamiltonian that depends on time smoothly. 
    Although such Floquet models are generally believed to obey Floquet eigenstate thermalization hypothesis~\cite{Lazarides14,DAlessio2014,Kim_ETH,Mori2018}, our construction leads to Floquet models that violate these expectations. 
    The Floquet model constructed from our approach is expected to have at least one eigenstate that is given by the MPS in the infinite range limit, where the Hermitian tangent space generator becomes exact. 
    We numerically verified that the Floquet spectra observed at finite sizes and ranges are indeed consistent with these expectations.

    The construction of MPS tangent space generators in this work is of both practical and conceptual utility. 
    Practically, our construction can be used to understand the most important operator terms needed for the optimal control along a specific entangled trajectory. 
    Although our construction generally returns a combination of all allowed operators within a certain range, in practice only a small number of operators has a dominant contribution. 
    It would be interesting to check whether the dominant operators agree with the expectations from counter-diabatic terms~\cite{Claeys_2019}, or if our approach is able to uncover different families of relevant operators. 
    In a different setting where one considers an MPS approximation to the ground state of a gapped Hamiltonian, the variational ansatz for quasiparticle excitations belongs to the tangent space~\cite{Laurens19,Haegeman,Zauner_Stauber_2018,Haegeman_2016,Haegeman_2013}. 
    Thus, exact tangent space generators constructed in this work may be used to excite quasiparticles above such gapped ground states. 
    
    Conceptually, our construction opens the door to the study of the properties of tangent space generators more systematically. 
    It is desirable to extend the relation between the structure of the tangent space generators and their complexity on the one hand, and the properties of the local MPS and the chosen tangent space direction on the other hand.

    Another intriguing question revealed by our work concerns the existence of \emph{finite-range} exact Hermitian tangent space generators. 
    We know simple examples of the exact MPS dynamics generated by operators that are sum of mutually commuting terms~\cite{De_Nicola_2021}. 
    However, it is desirable to understand if such a construction can be generalized beyond the sums of mutually commuting operators to ``frustration-free'' local generators of exact MPS dynamics (in analogy to the frustration free Hamiltonian construction)~\cite{Fannes1992,Fern_ndez_Gonz_lez_2014}. 
    If successful, these models may become instrumental in our understanding of unitary dynamics. 

    A different stride of our work is the establishment of the connection between generators of the tangent space in MPS manifolds and quantum many-body scars~\cite{Turner2017}. 
    Our results imply that any periodic trajectory over an MPS manifold has a parent Floquet model, whose dynamics is generated by a quasi-local Hamiltonian which smoothly depends on time, and where the corresponding MPS state is an exact eigenstate. 
    This invites a systematic study of quantum scars in Floquet models, and suggests that they may be more common than previously anticipated. 
    Understanding the microscopic mechanism that protects the MPS eigenstate of such Floquet models from hybridizing with other eigenstates and thermalizing remains an interesting open question. 
    Likewise, it may be interesting to understand the classical dynamics generated by such Floquet models obtained after  a TDVP projection~\cite{wenwei18TDVPscar,Laurens19,Michailidis2020}, and study the construction in the case of trajectories that generate a large amount of entanglement. 

    Finally, it may be interesting to extend our approach beyond the manifold of MPS, in particular, considering other variational forms of quantum wave functions. 
    Intuitively, we expect that variational ansatzes which are capable of capturing the unitary dynamics of the system may allow for a similar local or quasi-local construction of the corresponding tangent space generators. 
    Thus, understanding the structure of the tangent space generators may provide a useful test for the utility of other variational ans\"atze, and may also be useful for understanding the complexity and practical routes of state preparation beyond matrix product states.

    \section*{Acknowledgments} 
    \label{sec:acknowledgements}

    We thank L.~Piroli, S.~Garratt, and A.~Moln\'ar for insightful discussions. 
    This research was funded in part by the European Research Council (ERC) under the European Union’s Horizon 2020 research and innovation programme (grant agreements No.~850899 and No.~863476), the Austrian Science Fund (FWF) (Grant DOIs \href{https://doi.org/10.55776/COE1}{10.55776/COE1}, \href{https://doi.org/10.55776/P36305}{10.55776/P36305}, and \href{https://doi.org/10.55776/F71}{10.55776/F71}), and the European Union (NextGenerationEU).
    This work was performed in part at the Aspen Center for Physics, which is supported by National Science Foundation grant PHY-2210452.
    This research was supported in part by grant NSF PHY-2309135 to the Kavli Institute for Theoretical Physics (KITP). 
    For open access purposes, the authors have applied a CC BY public copyright license to any author accepted manuscript version arising from this submission.

    \appendix    

    \section{Free parameters}
    \label{app:free_params}

    In this Appendix we describe the various free parameters of our driving operator $o_{1,r}$, the mechanisms behind their emergence, and the relations between them. 
    Specifically, we discuss the following free parameters below:
    \begin{enumerate}
        \item Free parameters $\alpha_i$ that describe the distribution of the derivative. These parameters are defined in Eq.~\eqref{eq:alpha}; they are used to bound the leakage in Sec.~\ref{sec:hermitian_operator} and to optimize the fidelity of the approximate Hermitian driving scheme in Sec.~\ref{sec:example_convergence}.
        \item Free parameters $c_{x,y}$ that stem from the lack of any restrictions on the action of $o_{1,r}$ to states in $\mathcal{B}_\psi^{(r)}$. These parameters are defined in Eq.~\eqref{eq:free_params}, and they are used to numerically obtain operator space convergence in Sec.~\ref{sec:example_convergence}.
        \item Free parameters that might arise due to the changing the gauge of the MPS, potentially in a time-dependent way. They are not discussed in the main text, however in Appendix~\ref{app:bound} we use time-independent gauge invariance to set a fixed gauge for the bound.
        \item Free parameters emergent from the so-called telescoping series summation rule~\cite{Molnar_2018,Laurens19,Cirac_2021,Yang_2023}. These are also not discussed in the main text.
    \end{enumerate}

    \subsection{Distribution of the derivative} \label{app:derivDist}
    
    Let us first discuss $\alpha_i$, which are already defined in the main text in Eq.~\eqref{eq:alpha}. 
    The choice of these parameters has a clear effect on the driving operator $o_{1,r}$ and as a result on the leakage of the Hermitian driving operator $h_{1,r}=o_{1,r}+o^\dagger_{1,r}$ when acting on the MPS state $|\psi\rangle$. 
    As such we consider the choice of these parameters relevant as they can have an impact on the performance of the obtained driving protocol. 
    As we will discuss below, these parameters can be viewed as a subset of those arising from the telescoping summation, which we discuss in Appendix~\ref{app:telescope} below.

    \subsection{Action on the complement space} \label{app:complement}

    Next, we can consider the parameters $c_{x,y}$ defined in Eq.~\eqref{eq:free_params}. 
    These parameters, in the most generic case have $x\in\{\mathcal{V}_p^{(r)}\}$ and $y\in\{{\mathcal{B}_\psi^{(r)}}\}$, running over the basis of the complete Hilbert space and the image complement, respectively. 
    However, throughout the main text we considered a \emph{subset} of these parameters, where we restricted $x\in\{\mathcal{B}_\psi^{(r)}\}$. 
    This restricted set of parameters was used to obtain convergence in operator space in Sec.~\ref{sec:example_convergence}. 
    We note that while the restricted set of free parameters $c_{x,y}$  has no effect on the time dynamics of the chosen MPS state $|\psi\rangle$ these parameters will affect the dynamics of other states. 
    Thus, varying these parameters one may essentially obtain different generators of dynamics (or Floquet propagators when considered over a periodic trajectory) which are guaranteed to share only the single MPS eigenstate (in the limit $r\to\infty$).  

    Next, we discuss the remaining parameters, $c_{x,y}$ with  $x\in\{{\mathcal{A}_\psi^{(r)}}\}$ and $y\in\{{\mathcal{B}_\psi^{(r)}}\}$. 
    These parameters were not considered throughout our work due to the particular way in which we construct the Hermitian generators. 
    In particular, the non-zero value of such parameters would immediately contribute to the leakage that is given by $o^\dagger_{1,r}|\psi\rangle$. 
    Hence, although this freedom is not used in our work, it may be useful for other constructions of Hermitian tangent space generators. 

    Finally, let us comment that the parameters $c_{x,y}$ are not related to parameters $\alpha_i$. 
    Indeed, $c_{x,y}$ emerge as an addition, orthogonal to the term with the left-inverse ($I_r$), due to the restriction $y\in\{{\mathcal{B}_\psi^{(r)}}\}$.
    Conversely, changing $\alpha_i$ concerns the redistribution of the derivative MPS tensors ($\partial A$), which are contracted with the left-inverse, and thus acts non-trivially only on states in ${\mathcal{A}_\psi^{(r)}}$. 
    A similar statement can be made for the gauge transformations and telescoping series which we discuss next. 
    Note that while this clearly shows the effects of these parameters are strictly different to the other groups of free parameters, when considering $O_r$, the same may not be true when considering the Hermitian generator $H_r$ and further investigation is needed to fully understand the potential relations in that setting. 

    \subsection{Choice of MPS gauge} \label{app:gauge}

    Next we consider time-independent gauge transformations of the MPS $A'_i=G_{i-1}A_iG_{i}^{-1}$, with $G$ being some invertible $\chi\times\chi$ complex matrix, which is constant along the entire trajectory. 
    As our construction of the driving operators $o_{i,i+r-1}$ is invariant under such gauge transformation this will not lead to any additional freedom. 

    However, adding a time-dependent gauge transformation $G_i\to G_i(t)$, which depends on the time, or equivalently position along the chosen trajectory, could impact the resulting Hermitian operator and give rise to additional free parameters.
    This follows from the transformation for the derivative $\partial_t A'_i(t)=\partial_t\left(G_{i-1}(t)A_i(t)G_{i}^{-1}(t)\right)$ which modifies the derivative tensors by additional terms of the form $[\left(\partial_tG(t)\right)G^{-1}(t),G(t)A(t)G^{-1}(t)]$. 
    Although these transformations are non-trivial, their effect was not considered in the current work, in part because of the potential adverse impact on the norm of the leakage, which could require more complicated numerical optimization. 
    Lastly, the free parameters resulting from such a time-dependent gauge transformation are a subset of the telescoping degrees of freedom discussed below. 

    \subsection{Telescoping series \label{app:telescope}}

\begin{figure}[t]
        \centering
        \includegraphics[width=\linewidth]{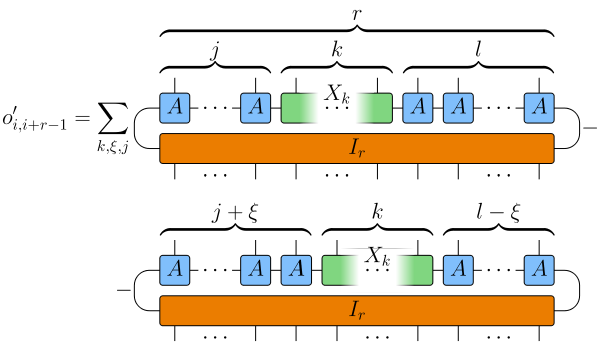}
        \caption{
        The action of the operators $o'_{i,i+r-1}$ (shown in the Figure) on the MPS state $|\psi(t)\rangle$ cancels out when summing over all sites, $\sum_i o'_{i,i+r-1} |\psi\rangle=0$. Therefore such operators can be added to the non-Hermitian driving operator without affecting the driving along the chosen MPS trajectory. The parameters $k,\xi,j$ in the Figure run over the following ranges: $k\in[1,r-1]$, $\xi\in[1,r-k]$, $j\in[0,r-k-\xi]$, and $l$ is defined as $l=r-k-j$. Finally, $X_k$ is an arbitrary tensor acting on $k$ sites.}
        \label{fig:telescoping}
    \end{figure}

    Finally, we briefly comment on the free parameters that can be obtained through the use of the telescoping sum. 
    Specifically, these are changes to the operators $o_{i,i+r-1}$, which do not impact $O_r|\psi\rangle=\sum_io_{i,i+r-1}|\psi\rangle$, as they should cancel out when summing over $i$. 
    A schematic construction of such operators $o'_{i,i+r-1}$ that can be added to our original construction ($o_{i,i+r-1}\to o_{i,i+r-1}+o'_{i,i+r-1}$) is presented in Fig.~\ref{fig:telescoping}. 

    Note that while these operators preserve the performance of the non-Hermitian driving scheme they can have an effect on the Hermitian scheme. 
    Indeed, while we do not consider this freedom explicitly in the manuscript, in order to avoid excess numerical optimization over the various parameters, this freedom is implicitly exploited when optimizing over $\alpha_i$.
    It is straightforward to see that the optimization over $\alpha_i$ fits into this broader family of telescoping freedom. 
    Furthermore, the time-dependent gauge deformations can also be seen as a subset of the same telescoping freedom. 
    In this sense, this telescoping freedom describes all deformations except for those described by $c_{x,y}$ (see Sec.~\ref{app:complement} above), which are clearly distinct when considering their effect on $o$.
    As mentioned previously this is easily seen by writing the deformations to our operator as $o^{\rm complement}_{i,i+r-1}=\sum_{x,y} c_{x,y}|x\rangle\langle y|$, where $\langle y|\in\mathcal{B}_\psi^{(r)}$, and $o^{\rm telescoping}_{i,i+r-1}=\sum_{x,y} d_{z,w}|z\rangle\langle w|$, where $\langle w|\in\mathcal{A}_\psi^{(r)}$ (see Sec.~\ref{sec:control_mps_sub_mps} for definitions of subspaces $\mathcal{A}_\psi^{(r)}$ and $\mathcal{B}_\psi^{(r)}$). 
    Since the subspace $\mathcal{B}_\psi^{(r)}$ is the orthogonal complement of $\mathcal{A}_\psi^{(r)}$, operators $o^{\rm complement}_{i,i+r-1}$ and $o^{\rm telescoping}_{i,i+r-1}$ are clearly distinct. 
    
    \begin{widetext}

    \section{The convergence bound}
    \label{app:bound} 
    
    In this Appendix we derive the convergence bound given in Eq.~\eqref{eq:bound}.
    From Eq.~\eqref{eq:locality}, we have that 
    \begin{equation}
            \lVert O_r^\dagger|\psi\rangle\rVert_2
            \leq
           \sum_i
            \lVert o_{i,i+r-1}^\dagger|\psi\rangle\rVert_2
    \end{equation}
    and thus, it suffices to bound the summands on the right hand side.
    We will consider the canonical $o_{i,i+r-1}$ constructed without the optimizations described in the main text, i.e., where the $\partial_t A^s$ sits only at the central site, $\alpha_i=\delta_{i,\lfloor\frac r 2\rfloor}$; any upper bound for this specific choice will clearly also upper bound $\lVert
    o_{i,i+r-1}^\dagger|\psi\rangle\rVert_2$ for an $o_{i,i+r-1}$ optimized over the derivative distribution parameters $\alpha_j$.

    In the following, we will assume an injective MPS. 
    Let us now consider 
    \begin{equation}
    \lVert o_{i,i+r-1}^\dagger|\psi\rangle\rVert_2^2
    =
    \langle  \psi |  o_{i,i+r-1} 
        o_{i,i+r-1}^\dagger|\psi\rangle\ ,
        \label{eq:appA_smallonorm}
    \end{equation}
    which corresponds to the tensor network shown in Fig.~\ref{fig:AppA_Fig1}a. 
    For simplicity we restrict to odd $r=2\ell+1$; for even $r$, one has two different $\ell$ on the left and right.
    We can now choose to work in any gauge we want, with one caveat: A gauge change $A^s\mapsto YA^s Y^{-1}$ will leave the diagram in Fig.~\ref{fig:AppA_Fig1}a unchanged (with $I_r$ defined in terms of the new $A$), except that  $\partial_t A^s$ is transformed to $Y(\partial_t A^s) Y^{-1}=:D_t A^s$ rather than to $\partial_t (YA^s Y^{-1})$ (which does make a difference since the gauge choice will typically depend on $t$; for the same reason, the outcome of the construction of $O_r$ will depend on the gauge). 
    We will denote this transformed derivative by $D_t$ in the following, and choose to work in the right-canonical gauge. 
    That is, the right and left transfer matrix fixed points will be $R=\id$ and $L\ge0$ with $\mathrm{tr}\,L=1$.

    Let us now simplify and rewrite the diagram in Fig.~\ref{fig:AppA_Fig1}a. 
    First, we define $\rho:=T^k_v$ to be the transfer matrix of $k$ sites read in the vertical direction. 
    Due to the construction of $I_r$ (see Eq.~\eqref{eq:inverseDef} in Sec.~\ref{sec:control_mps_sub_mps}), the $T^r$ in the middle cancels with one of the $(T^r_v)^{-1}$. 
    We are thus left with the diagram shown in Fig.~\ref{fig:AppA_Fig1}b, where the arrows indicate the direction in which the operators act. 
    Importantly, we interpret $\sigma^{-1}$, $L$, $R$, as well as the $A_i$ and $D_t A_i$ as vectors. 
    The expression in Fig.~\ref{fig:AppA_Fig1}b thus amounts to the formula
    \begin{subequations}
        \begin{align}
        \lVert o_{i,i+r-1}^\dagger|\psi\rangle\rVert_2^2
        &= 
        \sum_{s,s'}
        \big[(\sigma^{-1}|\otimes(D_tA^s|\otimes(\overline{D_tA^{s'}}|\big]
        \,
        \mathcal{P}^{-1}(\rho\otimes\rho\otimes \rho^T\otimes \rho^T)\mathcal{P}
        \,
        \big[|L)\otimes |R)\otimes |A^{s})\otimes |\overline{A^{s'}})\big]
        \\
        & \hspace*{-3em}= \mathrm{tr}\Bigg[
            \underbrace{
            \mathcal{P}\bigg\{
            \Big(\sum_s |A^{s})(D_tA^{s}|\Big)
            \otimes\Big(\,\overline{\sum_{s'} |A^{s'})(D_tA^{s'}}|\Big) \otimes
            \Big(\big[|L)\otimes |R)\big](\sigma^{-1}|\Big) 
        \bigg\}\mathcal{P}^{-1}}_{\displaystyle=:S}
        (\rho\otimes\rho\otimes \rho^T\otimes \rho^T)
        \Bigg]
        \label{eq:appBnd:trace-S-rho}
        \end{align}
    \end{subequations}
    Here $\mathcal{P}$ is a permutation of components in the tensor product which ensures that the expression matches that presented in the right hand side of the expression in Fig.~\ref{fig:AppA_Fig1}. 

    \begin{figure}[t]
        \includegraphics[width=0.85\linewidth]{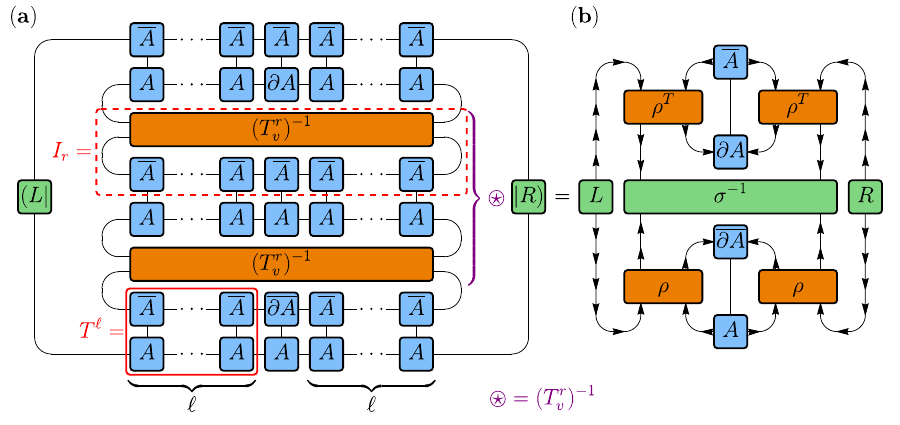}
        \caption{Diagrammatic representation of the norm from Eq.~\eqref{eq:appA_smallonorm} (left) and the simplified expression in Eq.~\eqref{eq:appBnd:trace-S-rho} (right). 
        Note that we consider the operator range to be odd $r=2\ell+1$, in the case of even operator range $r=2\ell$, one instead has $\ell$ and $\ell-1$ sites on the left and right sides of the derivative term respectively. 
        Note that the middle section (marked with $\protect\ostar$) simplifies to a single inverse, since it can be rewritten as simply $(T_v^r)^{-1}T_v^r(T_v^r)^{-1}$.
        On the right hand side the objects with all arrows pointing in or out (e.g. $\sigma^{-1}$, $L$, $R$) can be viewed as vectors (ket and bra respectively), while $\rho$ and $\rho^T$ are operators that act on their respective spaces. }
        \label{fig:AppA_Fig1}
    \end{figure}

    We now use that $T$ converges to $|R)(L|$. 
    From Ref.~\cite[Theorem 4.3 and Section 4.3]{SzehrReebWolf} (note that our $T$ is trace preserving due to the gauge choice), we have that 
    \begin{equation}
        \label{eq:appa:cbnorm_bound}
        \|T^n - |R)(L|\|_\diamond  
            \equiv \sup_{\omega\ge0,\,\|\omega\|_1=1}
                \big\|\big(T^n\otimes \id\big)(\omega)-\big(|R)(L|\otimes \id\big)(\omega)\big\|_1 
            \le C(\chi,|\lambda_2|) |\lambda_2|^n n^{\chi^2-1}\ ,
    \end{equation}
    with $\lambda_2$ the subleading eigenvalue (i.e. second largest in absolute value) of $T$, and where
    \begin{equation}
        C(\chi,|\lambda_2|)=4e^2\chi(\chi^2+1)\left(\frac{2}{1-|\lambda_2|}\right)^{3/2}
        \left(\mathrm{max}\big(1,\tfrac{1-|\lambda_2|^2}{|\lambda_2|}\big)\right)^{\chi^2-1}\ .
    \end{equation}
    Here, $\lVert\cdot\rVert_1$ is the trace norm and $\lVert\cdot\rVert_\diamond$ the diamond norm~\cite{aharonov1998,Watrous2018}. 
    By setting $\omega=\sum\lvert{i,i}\rangle\langle j,j\rvert$ (i.e., the unnormalized maximally entangled state) in Eq.~\eqref{eq:appa:cbnorm_bound} and noting that the $\|\cdot\|_1$ norm acts in the vertical direction (i.e., on the density matrix which is output by the channel)---which means that  $(T^n\otimes \id)(\omega)\equiv T^n_v$, $(|L)(R|\otimes \id)(\omega)\equiv L\otimes R$---we obtain that
    \begin{equation}
        \|\rho-\rho_\infty\|_1\le \chi\, C(\chi,|\lambda_2|)\,|\lambda_2|^\ell\, \ell^{\chi^2-1} =:\varepsilon(\ell)\ ,
    \end{equation}
    where the extra factor of $\chi$ comes from $\|\omega\|_1=\chi$, and we have defined $\rho_\infty = L\otimes R$.

    We will now use Eq.~\eqref{eq:orthogonality}, which states that the expectation value of a single $\partial_t A$ in the fixed point vanishes. 
    In our scenario, this means that the diagram in Fig.~\ref{fig:AppA_Fig1}b will be zero if we replace either both $\rho$ by $\rho_\infty$, or correspondingly both $\rho^T$.
    Note that this also works with $D_tA$, as can be seen by returning to the original gauge.
    We can now bound Eq.~\eqref{eq:appBnd:trace-S-rho} as follows:
    \begin{subequations}
        \begin{align}
        \mathrm{tr}\big[S(\rho\otimes \rho\otimes \rho^T\otimes \rho^T)\big]
            &=
            \mathrm{tr}\Big[S\:
                \big[(\rho\otimes\rho-\rho_\infty\otimes\rho_\infty) + \rho_\infty\otimes\rho_\infty \big]
                \otimes
                \big[(\rho^T\otimes\rho^T-\rho^T_\infty\otimes\rho^T_\infty) + \rho^T_\infty\otimes\rho^T_\infty \big]
            \Big]
        \\
        & \stackrel{(*)}{\le}
            \Big|\mathrm{tr}\big[S\:
                (\rho\otimes\rho-\rho_\infty\otimes\rho_\infty)
                \otimes
                (\rho^T\otimes\rho^T-\rho^T_\infty\otimes\rho^T_\infty) 
            \big]\Big|
            + 
        \nonumber\\
        &\hspace*{10em}+ 
            \underbrace{
            \Big|\mathrm{tr}\big[S\:
                \rho_\infty\otimes\rho_\infty
                \otimes
                (\rho^T\otimes\rho^T-\rho^T_\infty\otimes\rho^T_\infty) 
            \big]\Big|
            + \dots}_{\displaystyle \equiv 0}
            \label{eq:appBnd:vanishing}
        \\
        & \le
        \|S\|_\infty\:
        \|\rho\otimes\rho-\rho_\infty\otimes\rho_\infty\|_1^2\ ,
        \label{eq:appBnd:S-and-rho}
        \end{align}
    \end{subequations}
    where we have used the triangle inequality in $(*)$, and the terms in Eq.~\eqref{eq:appBnd:vanishing} vanish for the reason just discussed.
    We can further bound
    \begin{subequations}
        \begin{align}
        \|\rho\otimes\rho-\rho_\infty\otimes\rho_\infty\|_1
        &\le 
        \|\rho\otimes\rho-\rho\otimes\rho_\infty\|_1 + 
        \|\rho\otimes\rho_\infty-\rho_\infty\otimes\rho_\infty\|_1
        \\
        &=
        \|\rho\|_1\,\|\rho-\rho_\infty\|_1 + 
        \|\rho_\infty\|_1\,\|\rho-\rho_\infty\|_1 
        \\
        &\le 2 \chi\,\varepsilon(\ell)\ .
        \end{align}
    \end{subequations}

    It remains to bound $\|S\|_\infty$ (the spectral norm of $S$). 
    To this end, we define 
    \begin{equation}
     \beta :=
        \Big\|\sum_s |A^s)(D_tA^s|\Big\|_\infty\ ,
    \end{equation}
    which is a property of the chosen MPS trajectory. 
    In particular, it is bounded as long as rate of change $\partial_tA$ is bounded.
    Then, from Eq.~\eqref{eq:appBnd:trace-S-rho} we have
    \begin{equation}
        \|S\|_\infty = \beta^2\:\Big\| \big(|L)\otimes |R)\big)(\sigma^{-1}|\,\Big\|_\infty
       = \beta^2\,\sqrt{(L|L)\,(R|R)\,(\sigma^{-1}|\sigma^{-1})} 
        \ .
        \label{eq:appBnd:Sbnd-pre}
    \end{equation}
    We have that $(L|L)=\mathrm{tr}\,L^2\le 1$ (as $0\le L\le \id$ and $\mathrm{tr}\,L=1$) and $\mathrm{tr}\,R = \mathrm{tr}\,\id=\chi$. 
    To bound $(\sigma^{-1}|\sigma^{-1})$, where $\sigma=T^r_v$, denote by $\kappa:=|\lambda_\mathrm{min}(L)|$ the smallest eigenvalue of $L$, and choose $r$ such that $\varepsilon(r)\le \kappa/2$. 
    Then, $\|L\otimes R-\sigma\|_1\le \varepsilon(r) \le \kappa/2$, and thus $|\lambda_{\mathrm{min}}(\sigma)|\ge \kappa/2$. 
    Thus, 
    \begin{equation}
        (\sigma^{-1}|\sigma^{-1}) = \sum_i |\lambda_i(\sigma^{-1})|^2 = 
        \sum_i \frac{1}{|\lambda_i(\sigma)|^2} \le \frac{4}{\kappa^2}\chi\ .
    \end{equation}
    By inserting this in Eq.~\eqref{eq:appBnd:Sbnd-pre}, we arrive at $\|S\|_\infty \le {2\beta^2\chi}/{\kappa}$.
    
    We can now put this back into Eq.~\eqref{eq:appBnd:S-and-rho} to arrive at
    \begin{subequations}
        \begin{align}
            \lVert o_{i,i+r-1}^\dagger|\psi\rangle\rVert_2^2
            \le 
                 \frac{2\beta^2\chi}{\kappa}\,\big(2\chi\,\varepsilon(\ell)\big)^2 
                &= 
                 \frac{8\beta^2\chi^5}{\kappa}\,
                C(\chi,|\lambda_2|)\, |\lambda_2|^{2\ell}\, \ell^{\chi^2-1}\ ,
            \\
            &=
                 \frac{8\beta^2\chi^5\,C(\chi,|\lambda_2|)}{\kappa|\lambda_2|} 
                 \, |\lambda_2|^{r}\, \left(\frac{r-1}{2}\right)^{\chi^2-1}\ ,
        \end{align}
    \end{subequations}
    which proves the exponential decay of $o_{i,i+r-1}^\dagger|\psi\rangle$, and thus convergence to hermiticity, as $r$ is increased. 
    In particular, the right hand side is bounded by any exponential decay with a basis strictly larger than $|\lambda_2|$.
    Note that the obtained bound is uniform, as long as the conditioning number $\kappa$ of the entanglement spectrum is bounded from below along the trajectory, the correlation length does not diverge (i.e., $|\lambda_2|$ stays bounded away from $1$), and the rate of change of the MPS $\beta$ is bounded, which is always true for a periodic continuously differentiable trajectory (for the specific trajectory used as an example in our work, $\beta$ takes values between 2 and 14).

    \end{widetext}

    \section{Closed loop trajectory}
    \label{app:trajectory}

    \begin{figure}[b]
        \includegraphics[width=\linewidth]{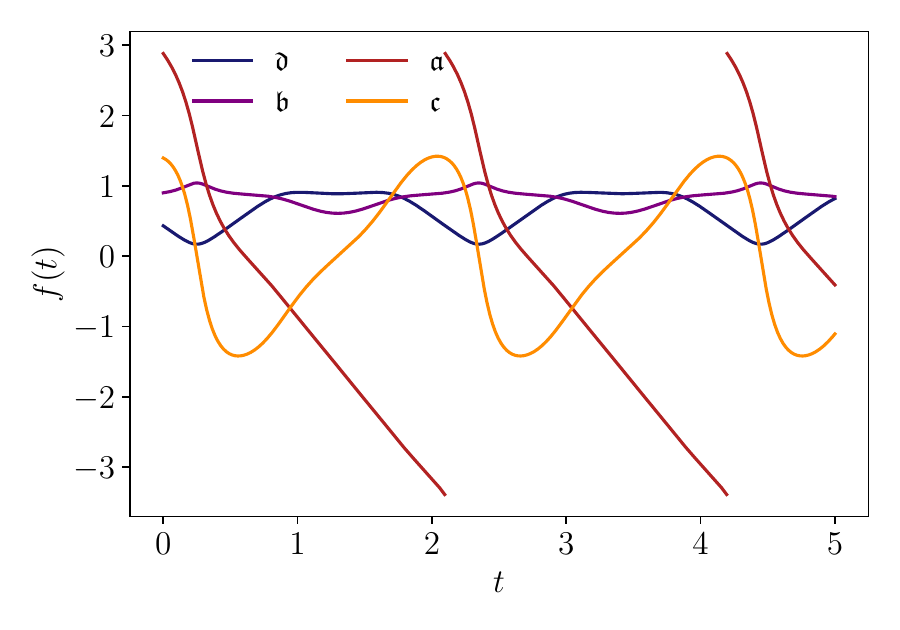}
        \caption{Time dependence of the MPS parameters as defined in Ref.~\cite{Michailidis2020,Ljubotina2022}. The trajectory was obtained by applying TDVP to an Ising Hamiltonian on the MPS in Eq.~\eqref{eq:isingmps} with the starting point set to $(\mathfrak{a},\mathfrak{b},\mathfrak{c},\mathfrak{d})=(0.2607,0.9,4.888,0.4308)$. }
        \label{fig:mps_params}
    \end{figure}

    In this Appendix we provide a more detailed description of the closed loop trajectory used in the main text. 
    The trajectory within the MPS manifold in Eq.~\eqref{eq:isingmps} was obtained in Ref.~\cite{Michailidis2020} using the TDVP projection of the unitary dynamics generated by the Ising model, however, this trajectory had large leakage, reflecting the rapid growth of entanglement in the dynamics generated by the Ising model. 
    The large leakage means that quantum dynamics generated by the Ising model did not follow the MPS trajectory, but would ``leak'' to other states outside of this MPS manifold. 
    The trajectory was studied in a previous control paper~\cite{Ljubotina2022}, where a variational approach was used to reduce the leakage along the trajectory. 
    Nevertheless, due to the variational character of the approach in Ref.~\cite{Ljubotina2022}, the reduced leakage remained non-zero.
    
    We note, that existence of periodic trajectories in the TDVP projection of unitary quantum dynamics is natural from the point of view of dynamical systems. 
    The TDVP projection turns the Schr\"odinger equation into a first order system of differential equations with a symplectic structure~\cite{Laurens19,Michailidis2020}. 
    In the typical case we expect that the resulting classical dynamics is non-integrable, and thus, as a chaotic system, it is expected to have an \emph{infinite} number of periodic trajectories. 

    Due to the fact that the projected dynamics only approximately capture the dynamics of the quantum system, the shortest periodic trajectories are typically the most relevant. 
    In particular, Ref.~\cite{Michailidis2020} found a relatively short periodic trajectory, with a period of $t_0\approx2.098$ in units where the leading coupling in the Hamiltonian was set to one. 
    The trajectory is defined by the time-dependent values of the MPS parameters $\mathfrak{a},\mathfrak{b},\mathfrak{c},\mathfrak{d}$, which are shown in Fig.~\ref{fig:mps_params}. 

    \section{Numerical study of Floquet unitary}
    \label{app:numerics}

    In this Appendix we present further numerical results on the properties of the Floquet propagator obtained by considering the driving along the trajectory, defined in Eq.~\eqref{eq:floquet}. 
    By construction, the Hamiltonian is translation invariant, thus we can resolve this symmetry and work in a reduced Hilbert space defined by the chosen momentum sector. 
    As our initial state lies in the zero momentum sector, we restrict our calculations to that sector exclusively. 
    In particular, we will be considering the properties of the spectrum of the Floquet propagator
    \begin{equation}
        U_F|\Phi_n\rangle=\phi_n|\Phi_n\rangle,
        \label{eq:app_fham}
    \end{equation}
    where $|\Phi_n\rangle$ are the eigenvectors and $\phi_n$ are the corresponding eigenvalues.
    The latter are related to those of the Floquet Hamiltonian $H_F=i\ln U_F$ as
    \begin{equation}
        e_n=i\ln\phi_n.
        \label{eq:app_eigvals}
    \end{equation}

    \begin{figure}[t!]
        \centering
        \includegraphics[width=1\linewidth]{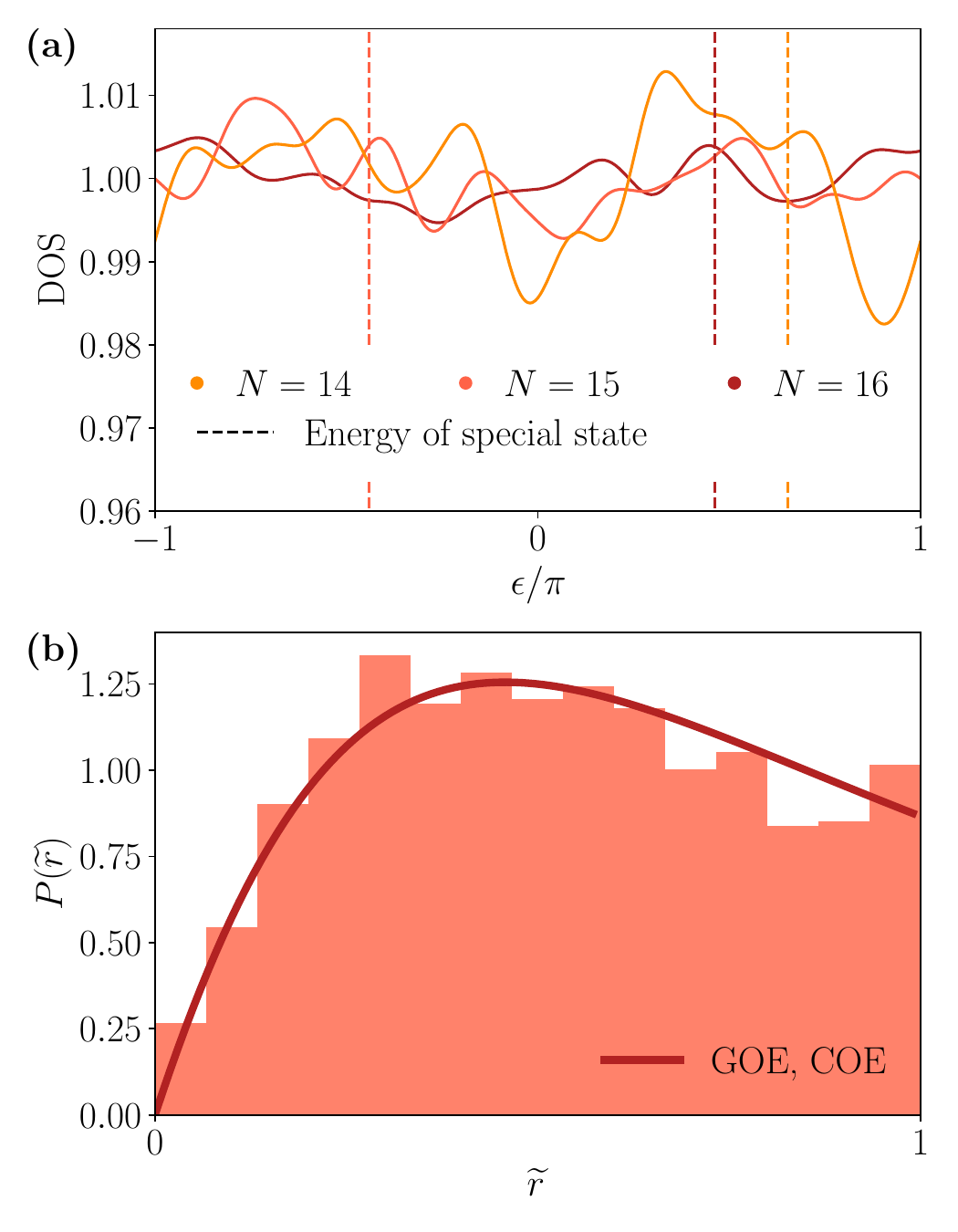}
        \caption{
            (a) Smoothed density of states (sDOS) with $\sigma^2 = 0.05)$ for different system sizes.
            (b) Spectral ratio distribution with $r=4$ and $N=14$. 
            The sDOS shows clear agreement with GOE predictions, as is expected for chaotic systems. 
            Furthermore, the sDOS data show no gap around the energy of our special state (eigenstate with the largest overlap with the MPS state -- energy marked with a dashed vertical lines), which can thus be viewed to be in the middle of the spectrum with no effective gaps closing with system size. 
        }
        \label{fig:numerics_level_spacing_dos}
    \end{figure}

    To construct the propagator, we use the QuSpin package~\cite{weinberg2017quspin,weinberg2019quspin} to build the driving Hamiltonian and then apply the \texttt{odientw} solver from SciPy~\cite{github} to find the solution of the time-dependent Schr\"odinger equation and get the propagator $U_t$ at each time step. 
    The solver was used with a maximum number of steps $10^7$ and an absolute error tolerance of $\texttt{atol}= 10^{-9}$.

    First, we consider the smoothed density of states (sDOS) of the propagator to determine that there are no unusual features in the vicinity of the energy of the MPS-like eigenstate of $U_F$. 
    We define the sDOS as follows:
    \begin{equation}
        \rho_{\sigma}(\epsilon) = \frac1D
        \sum_n \frac{1}{\sqrt{2\pi\sigma^2}} e^{-{(\epsilon-e_n)^2}/{2\sigma^2}},
    \end{equation}
    where $e_n$ are the eigenvalues of the Floquet Hamiltonian, $D$ is the dimension of the Hilbert space in the considered symmetry sector, and $\sigma$ is the smoothing parameter. 

    \begin{figure}[b]
        \centering
        \includegraphics[width=1\linewidth]{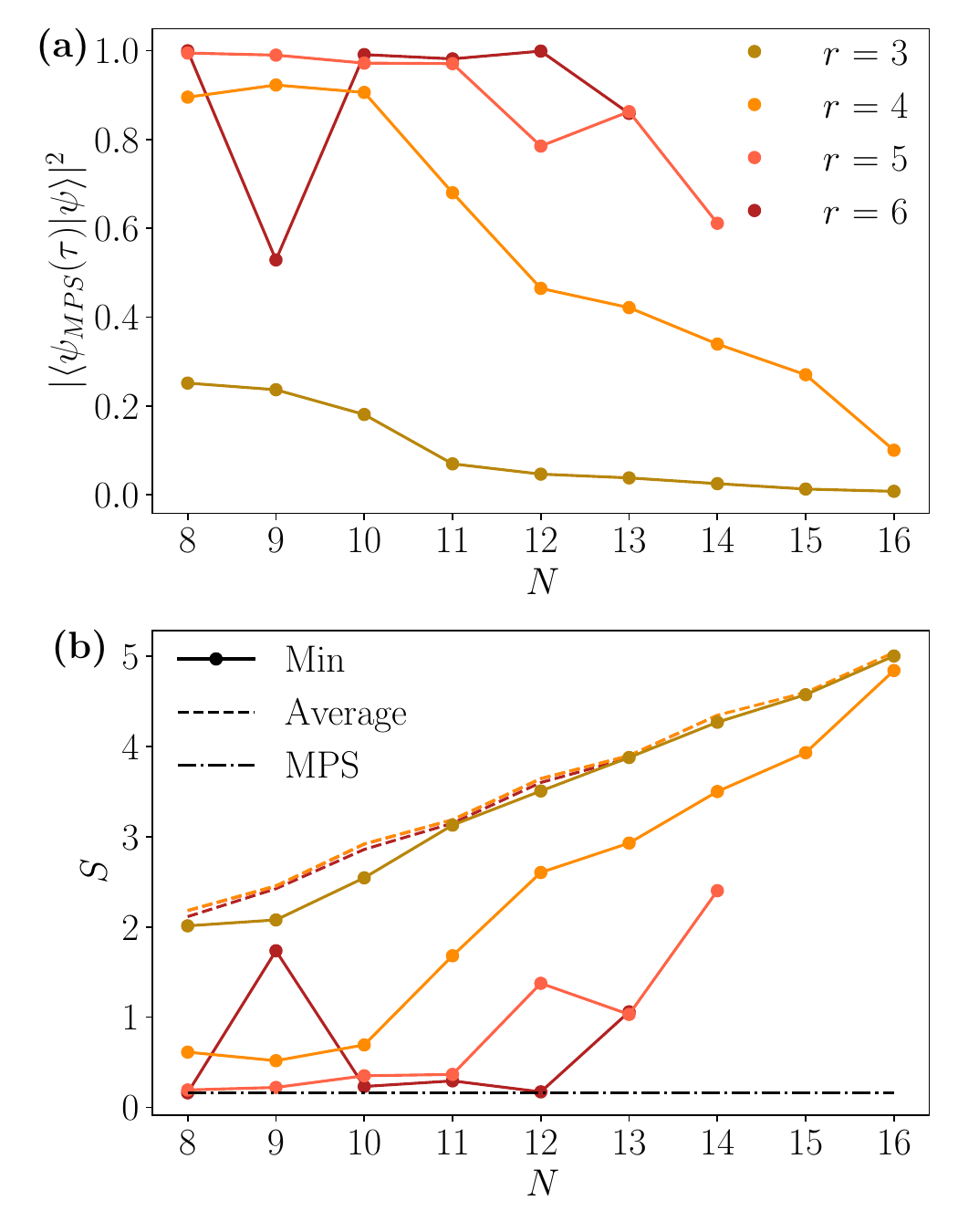}
        \caption{
            (a) Dependence of the maximal overlap of the initial state $\ket{\psi(0)}$ and eigenstates of $U(\tau)$ for different system sizes and operators. 
            (b) Entanglement entropy for different system sizes and operators. 
            For $r=4-6$ the eigenstates states that have the biggest overlap with the initial state and minimal entropy are the same.  
        }
        \label{fig:numerics_overlap_entanglement_expectation}
    \end{figure}
     
    In Fig.~\ref{fig:numerics_level_spacing_dos}(a) we observe that the sDOS has a constant value, although for varying system sizes some deviations exist, which reduce with the system size. 
    This suggests that the system is not far from the thermodynamic limit, where the sDOS is expected to be constant. 
    The oscillations observed in the sDOS also show no sign of correlation with the quasi-energy of the MPS-like eigenstate, which suggests that this eigenstate is not protected by a gap and can be viewed as a true middle-of-the-spectrum state.

    \begin{figure}[t]
        \centering
        \includegraphics[width=1\linewidth]{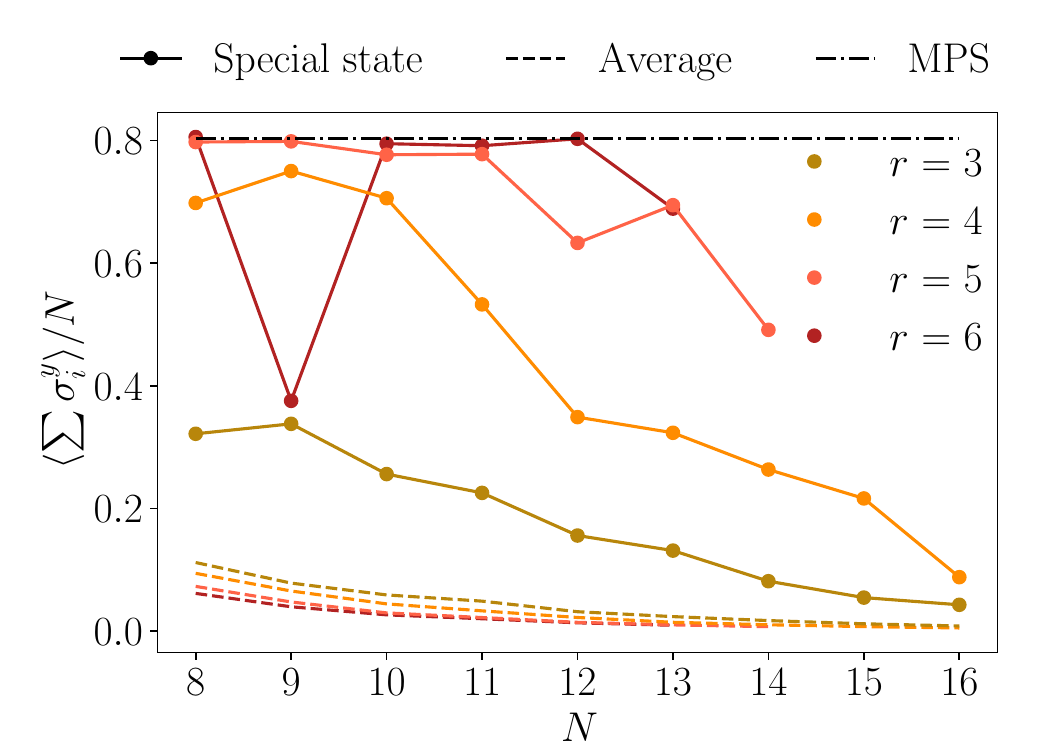}
        \caption{
            We show observable values for $\braket{\sum_i \sigma^y_i}/N$ on different eigenstates of the Floquet propagator for different supports $r$. 
            For $r=6$ the special eigenstate almost achieves the value of the initial MPS state, in line with expectations from the fidelity with the MPS state. 
            The averages over all eigenstates have significantly different values indicating that the special states violate Floquet ETH.
        }
        \label{fig:numerics_overlap_entanglement_expectation_2}
    \end{figure}

    Second, we study the distribution of the neighboring eigenvalue ratios $\tilde{r} = \frac{\mathop{\rm min}(s_n, s_{n-1})}{\mathop{\rm max}(s_n, s_{n-1})}$, where $s_n = e_{n+1}-e_{n}$, of the obtained propagator to determine whether it is generic or has any additional special features, apart from the single eigenstate we were aiming to obtain. 
    In Fig.~\ref{fig:numerics_level_spacing_dos}(b) we observe good agreement with random matrix theory (RMT) predictions \cite{atas2013distribution}, which aligns with expectations for chaotic models. 
    Specifically, the eigenvalues of the propagator match the results predicted by the circular orthogonal ensemble (COE).
    Interestingly COE results are usually associated with time reversal symmetry. 
    However this symmetry is not apparent in our example.
    The generator of the dynamics contains all possible Pauli terms and thus circular unitary ensemble (CUE) statistics would instead be expected.

    In case the constructed propagator leads to dynamics with no leakage, we expect to find the MPS state among its eigenstates. Fig.~\ref{fig:numerics_overlap_entanglement_expectation}(a) shows the maximal overlap of the desired MPS states with the eigenvectors of the evolution operator (or Floquet Hamiltonian).
    Our analysis reveals a notable trend, at small support (e.g. $r=3$) the overlap of the eigenstates with the initial MPS state shows no significant outliers.
    However, as the support is increased, the overlap shifts increasingly to a single state, which we shall from now on refer to as the \emph{special state}.
    This is in line with the expectation that at $r\to\infty$ the initial MPS state should be an exact eigenstate of the unitary. 
    Nevertheless, for finite support $r$, the estimate in Eq.~\eqref{eq:bound} suggests that the MPS state is not an eigenstate. 
    Furthermore, we observe that the drop in overlap appears later in system size for larger support, which is also consistent with the bound in Eq.~\eqref{eq:bound} which suggests perfect driving in the limit $r\to \infty$.
    
    Additionally, we examined the entanglement of these special states compared to the average entanglement across all eigenstates, shown in Fig.~\ref{fig:numerics_overlap_entanglement_expectation}(b). 
    Here $S=-\rm Tr \rho_A \ln \rho_A$ is the von Neumann entanglement entropy, where $\rho_A$ is the reduced density matrix of the left half of the system.
    In case of an odd system size the system is split such that $\lfloor N/2 \rfloor$ sites are in the left section. 
    The average entanglement is the same for operators with different supports and depends only on the system size, which is in line with expectations from Floquet ETH~\cite{Lazarides14,DAlessio2014,Kim_ETH,Mori2018}. 
    Conversely, the special state shows significantly lower entanglement. 
    By increasing the support of the driving Hamiltonian the entanglement of special states is reduced further and should in the limit $r\to\infty$ converge to the entanglement of the initial MPS state ($S_{\rm MPS}=0.1594$). 

    In a similar spirit, we studied the properties of the eigenstates of the evolution operator by examining the expectation values of local observables. 
    It is reasonable to expect that the state most similar to the MPS state would exhibit similar expectation values, while for the rest of the eigenstates one would expect Floquet ETH to hold. 
    In Fig.~\ref{fig:numerics_overlap_entanglement_expectation_2} we show the expectation values of $\sum_i \sigma^y_i/N$, which confirm this intuition. 
    Here $\sigma^\alpha$ are the corresponding Pauli matrices with $\alpha\in\{x,y,z\}$. 
    Special states for all support operators have significantly different expectation values of observables compared to the mean, signaling a violation of Floquet ETH for that state.
    Different operators such as $\sum_i \sigma^x_i/N$ and $\sum_i \sigma^z_i/N$ show qualitatively similar results and are thus not shown. 
    Furthermore, the expectation values of the special state tend towards the expectation values of the MPS state with increasing support $r$. 
    
    All our observations suggest that, within the predominantly chaotic spectrum of $U_F$ constructed from MPS tangent space generators, there exists a unique state exhibiting revivals, thereby violating the expectations set by the Floquet ETH~\cite{Lazarides14,DAlessio2014,Kim_ETH,Mori2018}. 
    Finally, we note that the jump systematically observed in Figs.~\ref{fig:numerics_overlap_entanglement_expectation} and \ref{fig:numerics_overlap_entanglement_expectation_2} for $r=6$ and $N=9$ appears to be an accidental hybridization event that can happen for finite $r$, where the MPS state is predominantly composed of two eigenstates with nearly identical energies. 

    \bibliography{bib}

\end{document}